\documentclass[manuscript]{aastex63}
\usepackage{graphicx}
\usepackage{supertabular}
\usepackage{longtable}
\usepackage{CJK}
\usepackage{tikz}
\usepackage{threeparttable}
\usepackage{verbatim}
\usetikzlibrary{shapes,arrows}
\received{----}
\revised{April 1, 2022}
\accepted{April 1, 2022}
\submitjournal{ApJS}

\begin{document}
\begin{CJK*}{UTF8}{gbsn}

\defcitealias{2022ApJS..259...12W}{Paper I}
\defcitealias{2019ApJ...886..108F}{F19}

\title{Dust Extinction Law in Nearby Star-Resolved Galaxies. II. M33 Traced by Supergiants}

\correspondingauthor{Jian Gao}
\email{jiangao@bnu.edu.cn; yiren@qlnu.edu.cn}

\author[0000-0003-3860-5286]{Yuxi Wang (王钰溪)}
\affiliation{Department of Astronomy, Beijing Normal University, Beijing 100875, People's Republic of China; \rm{\href{jiangao\@bnu.edu.cn}{jiangao@bnu.edu.cn}}}

\author[0000-0003-4195-0195]{Jian Gao (高健)}
\affiliation{Department of Astronomy, Beijing Normal University, Beijing 100875, People's Republic of China; \rm{\href{jiangao\@bnu.edu.cn}{jiangao@bnu.edu.cn}}}

\author[0000-0003-1218-8699]{Yi Ren (任逸)}
\affiliation{College of Physics and Electronic Engineering, Qilu Normal University, Jinan 250200, People's Republic of China; \rm{\href{yiren\@mail.bnu.edu.cn}{yiren@qlnu.edu.cn}}}

\author[0000-0003-2472-4903]{Bingqiu Chen (陈丙秋)}
\affiliation{South-Western Institute for Astronomy Research, Yunnan University, Kunming, 650500, People's Republic of China}



\begin{abstract}

The dust extinction curves toward individual sight lines in M33 are derived for the first time with a sample of reddened O-type and B-type supergiants obtained from the LGGS.
The observed photometric data are obtained from the LGGS, PS1 Survey, UKIRT, PHATTER Survey, GALEX, Swift/UVOT and XMM-SUSS.
We combine the intrinsic spectral energy distributions (SEDs) obtained from the ATLAS9 and Tlusty stellar model atmosphere extinguished by the model extinction curves from the silicate-graphite dust model to construct model SEDs.
The extinction traces are distributed along the arms in M33, and the derived extinction curves cover a wide range of shapes ($R_V \approx 2-6$), indicating the complexity of the interstellar environment and the inhomogeneous distribution of interstellar dust in M33.
The average extinction curve with $R_V \approx 3.39$ and dust size distribution $dn/da \sim a^{-3.45}{\rm exp}(-a/0.25)$ is similar to that of the MW but with a weaker 2175 $\,{\rm \AA}$ bump and a slightly steeper rise in the far-UV band.
%, implying that the overall interstellar environment in M33 resembles the diffuse region in the MW.
%The median value of the average dust size is $\overline{a} \approx 7.54$ nm, which is smaller than that of the MW.
The extinction in the $V$ band of M33 is up to 2 mag, with a median value of $ A_V \approx 0.43$ mag.
The multiband extinction values from the UV to IR bands are also predicted for M33, which will provide extinction corrections for future works.
The method adopted in this work is also applied to other star-resolved galaxies (NGC 6822 and WLM), but only a few extinction curves can be derived because of the limited observations.

\end{abstract}



\section{Introduction} \label{sec:intro}

Interstellar dust efficiently absorbs and scatters starlight, affecting observations and physical processes.
Dust extinction or dust attenuation is of vital importance to recover the intrinsic spectral energy distributions (SEDs) of celestial objects and infer the properties of dust.
Extinction represents the amount of light lost due to absorption and scattering of dust along a sight line.
The extinction at a given wavelength depends on the grain size distribution and the optical properties of the grains \citep{2020ARA&A..58..529S}.
In contrast to extinction, attenuation depends on both extinction and the complexity of star-dust geometry in galaxies, including scattering back into the sight line, varying column densities or optical depths and the contribution by unobscured stars \citep{2020ARA&A..58..529S}.

\citet[CCM hereafter]{1989ApJ...345..245C} found that the dust extinction law in the Milky Way (MW) from ultraviolet (UV) to near-infrared (IR) bands could be characterized by one parameter named the total-to-selective extinction ratio $R_V~[=A_V/E(B-V)]$, which depends on the interstellar environment along the sight line.
However, the CCM extinction law is limited to only a set of sight lines in the MW, and it is not generally applied to external galaxies \citep{2015ApJ...815...14C}.
The properties of dust extinction curves or dust attenuation curves in galaxies and the physical mechanisms that shape them are fundamental extragalactic astrophysics questions and are important for deriving the physical properties of galaxies \citep{2020ARA&A..58..529S}.
On the one hand, external galaxies allow us to study dust in diverse interstellar environments, which is a necessary intermediate step to understanding distant galaxies.
On the other hand, whereas interpretation can sometimes be difficult in the MW disk because we see the projected material of the entire disk, high-latitude observation of face-on galaxies can provide clearer sight lines \citep{2018ARAandA..56..673G}.

Recently, an increasing number of extinction or attenuation laws have been quantified in the Large Magellanic Cloud (LMC, \citealt{1978Natur.276..478N,1985ApJ...288..558C,1986AJ.....92.1068F,2003ApJ...594..279G}), the Small Magellanic Cloud (SMC, \citealt{1982A&A...113L..15L,1984A&A...132..389P,2003ApJ...594..279G}), M31 \citep{1996ApJ...471..203B,2014ApJ...785..136D,2015ApJ...815...14C,2022ApJS..259...12W} and M33 \citep{1999ApJ...519..165G,2017PhDT.......221H,2022AJ....163...16M}, showing variations in the extinction/attenuation curves and the complexity of interstellar environments in external galaxies.
Although the average extinctions in the LMC and M31 are similar to that in the MW, the extinction curve in the bar region of the SMC rises steeply in the UV bands and lacks 2175 $\,{\rm \AA}$.

For the late-type spiral M33 (Sc, \citealt{1973ugcg.book.....N}) ($\approx$ 840 kpc \citealt{1991ApJ...372..455F}), which is the third largest member in the Local Group Galaxies, the latest study on attenuation was carried out by \citet{2022AJ....163...16M}.
\citet{2022AJ....163...16M} combined archival images from UV to IR to derive the ages, masses, and the values of $E(B-V)$ for the young star cluster population in M33 and found that all the star clusters have moderate-to-small internal extinction [$E(B-V) < 0.6$ mag].
\citet{2017PhDT.......221H} imaged the galaxy from FUV to NIR and measured the spatial variation of the dust attenuation law in M33 for the first time.
They found that the attenuation curves tend to be steeper and with an MW-like 2175 $\,{\rm \AA}$ bump between the arms in M33, while along the arms, the curves seem to be shallower with a weak 2175 $\,{\rm \AA}$ bump.
The median attenuation curve derived in \citet{2017PhDT.......221H} is quite steep with a 2175 $\,{\rm \AA}$ bump and is somewhat different from the fairly shallow attenuation curve with a strong 2175 $\,{\rm \AA}$ bump obtained by \citet{1999ApJ...519..165G} in the M33 nucleus study.
\citet{2017PhDT.......221H} found a median value of extinction in $V$ band $A_V = 0.53$ mag, which is twice the value of the fairly small mean amount of dust extinction ($A_V \approx 0.25$ mag) derived from the star formation study in M33 by \citet{2009A&A...493..453V} because of the different assumed stellar models and the lack of FIR observations in \citet{2017PhDT.......221H}.
The dust attenuation laws derived in \citet{2017PhDT.......221H} and \citet{1999ApJ...519..165G} allow us to understand the effect of dust on light and analyze the dust properties on a large scale.
However, the extinction curve toward the individual sight lines in M33 has never been calculated before, which can provide us with a better, more detailed understanding of the properties and distribution of the dust.

With the improvement of the observation resolution, individual stars in M33 could be distinguished and their photometry information and spectral types could be obtained, providing us with a completely new prospect for exploring the extinction law toward individual sight lines in M33.
\citet[\citetalias{2022ApJS..259...12W} hereafter]{2022ApJS..259...12W} derived dozens of extinction curves toward individual sight lines in M31 with the combination of the intrinsic SEDs from the stellar model atmospheres and model extinction curves from the dust model.
In this work, the method adopted in \citetalias{2022ApJS..259...12W} is also applied to calculate the extinction curves in M33.
%For star-resolved galaxies, the pair method \citep{1970IAUS...36...28B} is extensively adopted to obtain the extinction law, which compares the spectrum of a reddened star with that of an unreddened (or slightly reddened) star with the same spectral type or with the intrinsic spectrum obtained from the stellar model atmosphere.
%Based on the traditional pair method,  proposed an improved pair method that combines the intrinsic SEDs from the stellar model atmospheres and the model extinction curves from the dust model and derived dozens of extinction curves toward individual sight lines in M31.
%In this work, the improved pair method is applied to M33.
We select the bright O-type and B-type supergiants in M33 from the Local Group Galaxies Survey (LGGS, \citealt{2016AJ....152...62M}) as the extinction tracers following \citetalias{2022ApJS..259...12W}.
Using the photometry available online, the spectral energy distribution (SED) for each tracer from UV to near-IR is constructed, the details of which are shown in Section \ref{sec:data}.
The method of forward modeling the SED to obtain the dust extinction law is described in Section \ref{sec:method}.
Section \ref{sec:re} presents the extinction curves derived in this work and the discussion.
Finally, our conclusions are summarized in Section \ref{sec:summary}.

% =========================================
\section{Data and sample} \label{sec:data}

\begin{figure}[ht!]
    \centering
	\includegraphics[scale = 0.5]{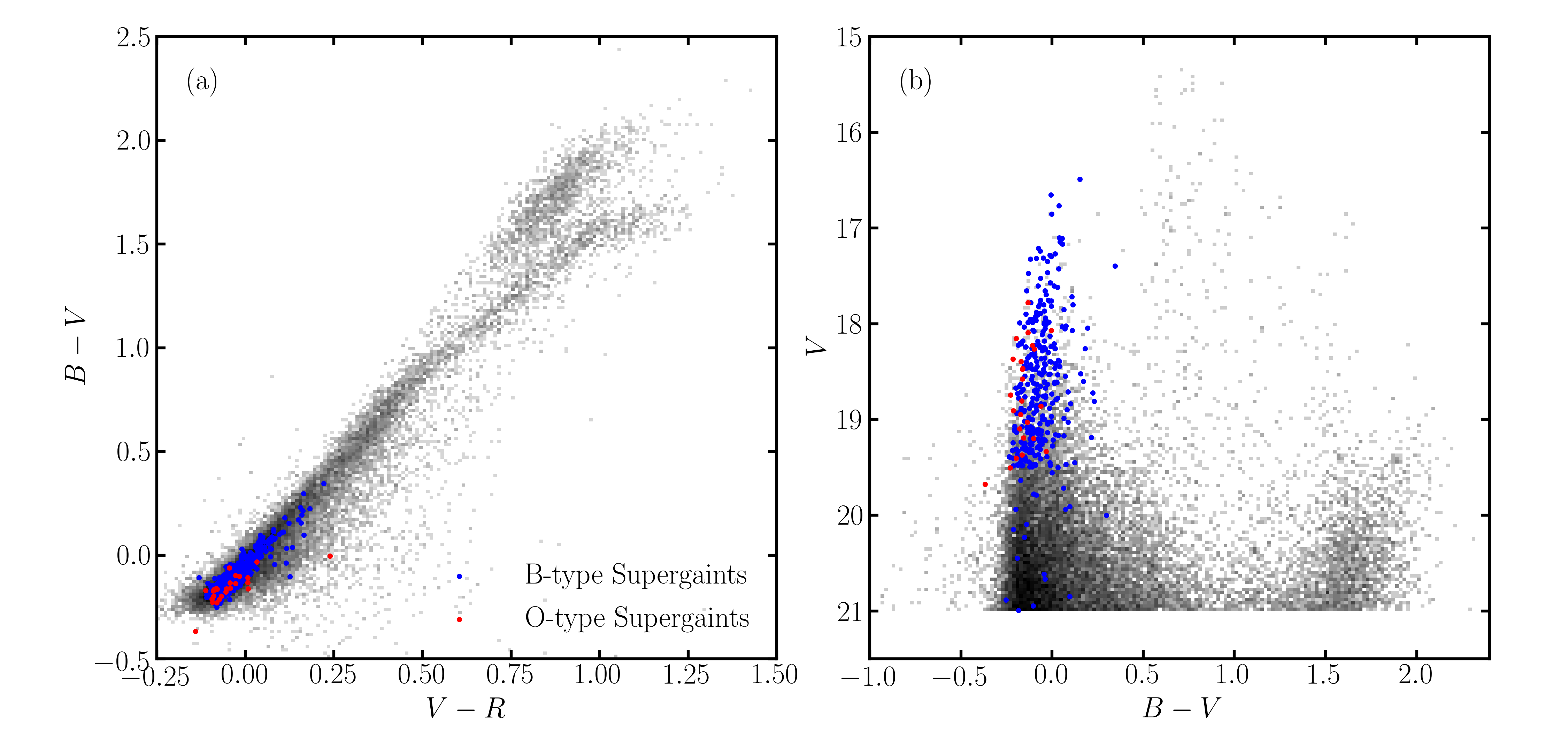}	
	\caption{Color-color diagrams (a) and color-magnitude diagrams (b) for all the LGGS sources (gray dot) and the selected O-type (red dot) and B-type (blue dot) supergiants.
	 \label{fig:samples} }
\end{figure}

As in \citetalias{2022ApJS..259...12W}, we selected the isolated O-type and B-type supergiants from the LGGS catalog \citep{2016AJ....152...62M} as the extinction tracers in M33 because supergiants are usually free of circumstellar dust and relatively bright
\citep{2018MNRAS.478.3467S,2019ApJS..241...32L}.
The LGGS catalog contains 146,622 stars in M33, of which 130 and 471 are confirmed to be O-type and B-type stars, respectively \citep{2016AJ....152...62M}.
In this work, the isolated O-type and B-type supergiants from the LGGS catalog are also selected as extinction tracers to explore the extinction law in M33.
25 O-type supergiants and 318 B-type supergiants constitute the extinction sample for M33 in this work.
Because of the limitation of ground-based telescopes, OB associations or binaries may be identified as single OB stars.
The optical images obtained from the Hubble Space Telescope (HST) in the $F475W$ and $V$ bands ($F547W$, $F555W$, $F569W$) are adopted to check the reliability of the extinction tracers.
There are 205 tracers in the extinction sample can be found in the HST/F475W image or the HST/V image, of which 200 stars seem to be single stars in the HST images.
While 5 sources are suspect because they overlap with other celestial objects and cannot be distinguished in the HST images.
As a result, we suggest that 98\% of the isolated supergiants from the LGGS catalog are reliable for calculating the dust extinction law in M33.
The $V-R/B-V$ diagram and $B-V/V$ diagram for all LGGS sources and the supergiants in the extinction sample are plotted in Figure \ref{fig:samples}.

We construct the observed SED for each tracer using the photometric data from the LGGS catalog \citep{2016AJ....152...62M} in the $U,~B, ~V,~R,~I$ bands, the United Kingdom Infrared Telescope (UKIRT, \citealt{2013ASSP...37..229I}) in the $J,~H,~K$ bands\footnote{The UKIRT $JHK$ brightness for the extinction tracers are processed by \citet{2021ApJ...907...18R}.}, the Panoramic Survey Telescope and Rapid Response System release 1 Survey (Pan-STARRS PS1, \citealt{2016arXiv161205560C}) in the $g,~r,~i,~z,~y$ bands, the XMM-Newton Serendipitous ultraviolet source survey (XMM-SUSS, \citealt{2019yCat.2356....0P}) in the $UVW2,~UVM2,~UVW1$ bands and the swift ultraviolet and optical telescope (Swift/UVOT, \citealt{2015yCat.2339....0Y}) in the $UVW2,~UVM2,~UVW1$ bands, as mentioned in \citetalias{2022ApJS..259...12W}.
The selection criteria for these catalogs are also the same as those in \citetalias{2022ApJS..259...12W}.

Instead of using the photometry from the Panchromatic Hubble Andromeda Treasury (PHAT) Survey \citep{2014ApJS..215....9W} adopted in \citetalias{2022ApJS..259...12W}, we obtain the photometry in M33 from the Panchromatic Hubble Andromeda Treasury: Triangulum Extended Region (PHATTER, \citealt{2021ApJS..253...53W}).
The PHATTER survey \citep{2021ApJS..253...53W} presents panchromatic resolved stellar photometry for 22 million stars in the Local Group dwarf spiral Triangulum (M33), derived from HST observations with the Advanced Camera for Surveys in the optical bands ($\lambda_{F475W} = 0.473~\mu {\rm m}$, $\lambda_{F814W} = 0.798~\mu {\rm m}$) and the Wide Field Camera 3 in the near-UV ($\lambda_{F275W} = 0.272~\mu {\rm m}$, $\lambda_{F336W} = 0.336~\mu {\rm m}$) and near-IR bands ($\lambda_{F110W} = 1.120~\mu {\rm m}$, $\lambda_{F160W} = 1.528~\mu {\rm m}$).
The survey covers $\sim$ 14 square kpc of the sky and extends to 3.5 kpc from the center of M33.
The PHATTER catalog is the largest stellar catalog for M33.
We check the GST (``good star") quality for each photometric point of each tracer and select the photometry with GST flag = `0', which means that the source passes the GST criteria in \citet{2021ApJS..253...53W}.
PHATTER photometry can thus be applied for 2 O-type supergiants and 36 B-type supergiants. 

In addition, we adopt UV data from the Galaxy Evolution Explorer (GALEX, \citealt{2017ApJS..230...24B}) in this work.
GALEX \citep{2005ApJ...619L...1M} performed the first sky-wide UV surveys with different coverage and depth \citep{2007ApJS..173..682M,2009Ap&SS.320...11B}, yielding observations in the following two broad bands: far-UV (FUV, $\lambda_{\rm eff} \approx 1528~\,{\rm \AA}$) and near-UV (NUV, $\lambda_{\rm eff} \approx~2310~\,{\rm \AA}$) \citep{2017ApJS..230...24B}.
Unfortunately, the M33 sky area is not completely covered by the observation of the GALEX.
In addition, it is estimated that most of the tracers in this work could be too faint for the GALEX with the typical depth of $m_{FUV} = 19.9$ mag and $m_{NUV} = 20.8$ mag to detect\footnote{Regardless of the dust extinction, the observed AB magnitudes in the $FUV$ and $NUV$ bands for a tracer with the median spectral type of B2 can be estimated with $T_{\rm eff} = 18000$ K, log$(g)$ = 2.50 and $d$ (distance) = 840 kpc.
The derived values of $m_{FUV} = 19.84$ mag and $m_{NUV} = 19.90$ mag plus the effect of dust extinction could be greater than the typical depth of the GALEX.}.
%We therefore suggest that most of the tracers in this work are too faint to be detected by the GALEX.
As a result, the UV data from the GALEX can be obtained for only a few tracers.
We first check the artifact flag and the extraction flag for each photometric point of the tracers and eliminate spurious sources.
We then eliminate the foreground photometry, which is too bright to fit the whole SED well.
Finally, the GALEX data for only 1 O-type supergiant and 2 B-type supergiants are retained.

Above all, at most 27 bands of photometric data from UV to near-IR are obtained for each star.
We summarize the selection criteria and the number of photometric points adopted for each catalog in Table \ref{Tab:data}.
%As mentioned in \citetalias{2021arXiv211109523W}, \emph{pcFlag} is also used in this work to show the coverage of passbands adopted in the calculation.
%\emph{U} in \emph{pcFlag} means the results are derived with UV data (here, this refers to the passbands bluer than the $U$ band), while \emph{V} and \emph{I} in \emph{pcFlag} are short for visual bands and near-IR bands (here, this refers to the passbands redder than the  $y$ band), respectively.

\newpage

\begin{deluxetable*}{cccccccccc}
	\tablecaption{Selection criteria and amount of photometric data adopted for each catalog \label{tab:data}}
		\tablehead{	
		\colhead{Catalog} & \colhead{Selection criteria $^b$} & \colhead{O-type supergiants} &  \colhead{B-type supergiants}   \label{Tab:data}
		}
	\startdata
		LGGS$^a$     & Supergiants; Cwd = `I'; V $<$ 21 mag & 25 & 318 \\
		\hline
		PS1          & Check information flags; Magnitude error $<$ 0.1 mag & 6 & 126 \\
		UKIRT/WFCAM  & $(J-H) < 0.3$ mag; $(H-K) < 0.16$ mag & 4  & 115  \\
		Swift/UVOT   & Extended flag = `0'; Quality flag = `0' & 5  & 34\\
					%& Eliminate foreground photometry & &  \\
		PHATTER Survey & ST flag = `0'; GST flag = `0' & 2 & 36 \\		
		XMM-SUSS     & Source quality flag = `FFFFFFFFFFFF'& 1  & 14  \\
		    %& Eliminate foreground photometry & & \\					
		GALEX        & Artifact flag = `0'; Extraction flag = `0' & 1  & 2
					%& Eliminate foreground photometry & &
	\enddata
	\tablecomments{$^a$ The LGGS catalog contains 130 O-type and 471 B-type stars in M33.
	The 25 selected O-type and 318 B-type isolated supergiants constitute the sample of extinction tracers in this work.\\
	$^b$ The criteria used to select the photometry from the LGGS, the UKIRT, the PS1 survey, the XMM-SUSS and the Swift/UVOT are the same as those in \citetalias{2022ApJS..259...12W}.
	The criteria for the UKIRT data refer to the results in \citet{2019ApJ...877..116W}.}
\end{deluxetable*}

% ======================================

\section{Method} \label{sec:method}

For star-resolved galaxies, the pair method \citep{1970IAUS...36...28B} is extensively adopted to obtain the extinction law, which compares the spectrum of a reddened star with that of an unreddened (or slightly reddened) star with the same spectral type.
In order to eliminate the influence of the limited unreddened standard stars and the mismatch error in the use of the pair method, \citet{2005AJ....130.1127F} proposed to use the stellar model atmospheres to derive the intrinsic SEDs rather than the unreddened standard stars.

Based on this ``extinction without standards'' technique, we first combine the intrinsic SED from the stellar model atmospheres extinguished by the model extinction curves to construct the model SEDs for the tracers, and then derive the extinction curves by fitting the model SEDs to the observed data.
Instead of the mathematical extinction models such as CCM \citep{1989ApJ...345..245C}, FM90 \citep{1990ApJS...72..163F} and F04 \citep{1999PASP..111...63F,2004ASPC..309...33F,2007ApJ...663..320F} extinction laws that are widely used in many works, the classic silicate-graphite dust model is adopted to model the dust extinction law as \citetalias{2022ApJS..259...12W}, so that the dust properties can also be analyzed besides obtaining the extinction curve.
In addition, the extinction curves derived from the dust model are more applicable in various interstellar environments than the parameterized extinction curves.

%In \citetalias{2021arXiv211109523W}, we proposed an improved pair method to probe the dust extinction law in M31, which is also applied in this work to calculate the extinction curves in M33.
%The improved pair method combined the intrinsic SEDs from the stellar model atmosphere extinguished by the model extinction curves to construct the model SEDs for the tracers.
%Instead of the mathematical extinction model, the classic silicate-graphite dust model is adopted to derive the dust extinction law so that the extinction curve and the dust properties can be obtained at the same time.
The detailed calculation process in this work can be found in Figure 3 of \citetalias{2022ApJS..259...12W}.
The construction of the model SEDs for M33 is described in detail in Section \ref{subsec:model_sed}, and Section \ref{subsec:fitting} describes the fitting of the model SEDs to the observed data.

\subsection{Model SEDs} \label{subsec:model_sed}

Theoretically, the observed SED $F^{\rm obs}_{\lambda}$ of a reddened star can be expressed as follows:
\begin{equation}
F^{\rm obs}_{\lambda}=F^{\rm int}_{\lambda}{\theta}_{R}^2 10^{-0.4A_{\lambda}},
\end{equation}
where $F^{\rm int}_{\lambda}$ is the intrinsic surface flux of the star at wavelength ${\lambda}$, ${\theta} \equiv (R/d)^2$ is the angular radius of the star (where $d$ is the distance and $R$ is the stellar radius), and $A_{\lambda}$ is the absolute extinction/attenuation of the stellar flux by intervening dust at $\lambda$ \citep{2005AJ....130.1127F}.

A Tlusty stellar model atmosphere \citep{2017arXiv170601859H,2017arXiv170601937H,2017arXiv170601935H} and an ATLAS9 stellar model atmosphere \citep{2003IAUS..210P.A20C} are adopted to obtain the intrinsic SEDs $F_{\lambda}^{\rm int}(T_{\rm eff}, {\log}~g, Z)$.
With 27 effective temperature values ($27500~{\rm K} \leq T_{\rm eff} \leq 40000~{K}$ with 2500 K steps for O-type supergiants, $10000~{\rm K} \leq T_{\rm eff} \leq 30000~{K}$ with 1000 K steps for B-type supergiants), 7 surface gravity values ($3.00 \leq {\rm log}~g \leq 3.50$ with 0.25 dex steps for O-type supergiants, $2.25 \leq {\rm log}~g \leq 3.00$ with 0.25 dex steps for B-type supergiants) and the solar value of the metallicity, a grid of intrinsic SEDs for M33 is constructed.

The model extinction curves in this work are also derived from the silicate-graphite dust model with the same exponential cutoff power-law grain size distribution proposed by Kim, Martin and Hendry (hereafter KMH, \citealt{1994ApJ...422..164K}) with $a_c$ fixed to 0.25 $\mu$m [$dn/da \sim a^{-\alpha}{\rm exp}(-a/0.25)$] for both components, as adopted in \citetalias{2022ApJS..259...12W}.
Detailed information on the chemical abundances of the interstellar environment in M33 is still controversial, and it is difficult to quantify the chemical abundances of the dust in M33.
However, some studies (e.g., \citealt{2007A&A...470..843M,2007A&A...470..865M,2016MNRAS.458.1866T,2019ApJS..241...35R,2021ApJ...907...18R}) suggest that the abundances in M33 are close to the protosolar values \citep{2009ARA&A..47..481A}, which are adopted in some recent works (e.g., \citealt{2017ApJ...841...20N,2021ApJ...908...87N,2021ApJ...907...18R}).
As a result, according to the previous works in other galaxies [e.g., MW \citep{2014P&SS..100...32W}, M31 (\citealt{2014ApJ...780..172D}; \citetalias{2022ApJS..259...12W}), NGC 4722 \citep{2020P&SS..18304627G}, etc], 
we assume that the interstellar abundances in M33 are similar to the protosolar values \citep{2009ARA&A..47..481A} and adopt a typical value of $f_{cs} = 0.3$ for the mass ratio of graphite to silicate, which means 
%the mass ratio of graphite to silicate $f_{cs} = 0.3$ means 
the elements of Fe, Mg and Si are all in the solid phase and constrained in silicate dust, and the fraction of gas-phase carbon is 50\%\footnote{In addition to fixing the mass ratio of graphite to silicate to $f_{cs} = 0.3$, we also adopt another typical value of $f_{cs} = 0.6$ \citep{2013EP&S...65.1127G,2014P&SS..100...32W,2015ApJ...807L..26G,2020P&SS..18304627G}, which means all graphite is in the solid phase.
However, it is found that the change of $f_{cs}$ does not substantially change the results.}.
We then derive a grid of the model extinction curves $A_{\lambda}(\alpha,A_V)$ with 56 values of $\alpha$ ($0.5 \leq \alpha \leq 6.0$ with 0.1 dex steps) and 51 values of $A_V$ ($0 \le A_V \le 5$ with 0.1 mag steps).

By combining the intrinsic SEDs and the model extinction curves, a large grid of monochromatic flux extinguished by dust can be derived as follows:
\begin{equation}
F_{\lambda}^{\rm mod}(\alpha, A_V, T_{\rm eff}, {\rm log}~g, Z)=F^{\rm int}_{\lambda}(T_{\rm eff}, {\rm log}~g, Z){\theta}_{R}^2 10^{-0.4A_{\lambda}(\alpha,A_V)},
\end{equation}
To compare the model SEDs with the observed photometry, the model band flux can be calculated as follows:
\begin{equation}
F^{\rm mod}_i = \frac{\int \lambda B_i(\lambda)F_{\lambda}^{\rm mod}d \lambda}{\int \lambda B_i(\lambda)d \lambda}
\end{equation}
where $B_i({\lambda})$ is the bandpass response function for the $i$th band.
The flux for each band is thus obtained from the response function and model SEDs, including the intrinsic SEDs and the model extinction curves.

% ==========

\subsection{Fitting Model SEDs to Observed Data} \label{subsec:fitting}

Since the tracers in M33 are paired with the stellar model atmospheres, the foreground MW dust extinction must be removed.
We adopt an MW foreground extinction component of $E(B-V) \approx 0.06$ mag \citep{2020ApJ...905L..20R}, assuming an $R_V = 3.1$ CCM dust, as a part of the fitting process.

The EMCEE fitting code \citep{2013PASP..125..306F} is used to fit the model SEDs to the observed data.
It is a Markov-Chain Monte Carlo (MCMC) ensemble sampler and helps obtain the most suitable parameters and the corresponding confidence intervals.
Gaussian likelihood is adopted, and flat priors are imposed on $\alpha$, log$(g)$ and $A_V$ in this work.
A Gaussian prior is imposed on the effective temperature log$(T_{\rm eff})$ based on the spectral type from the LGGS catalog \citep{2016AJ....152...62M} with one subclass in the spectral type considered the uncertainty (\citealt{2015ApJ...815...14C}; \citetalias{2022ApJS..259...12W}).
%The prior of the surface gravity log$(g)$ is based on the spectral type.
The calibration of the spectral type to log$(T_{\rm eff})$ and log$(g)$ refers to \citet{2000asqu.book.....C} and \citet{2008flhs.book.....C}.

\begin{figure}
	\centering
	\includegraphics[scale=0.4]{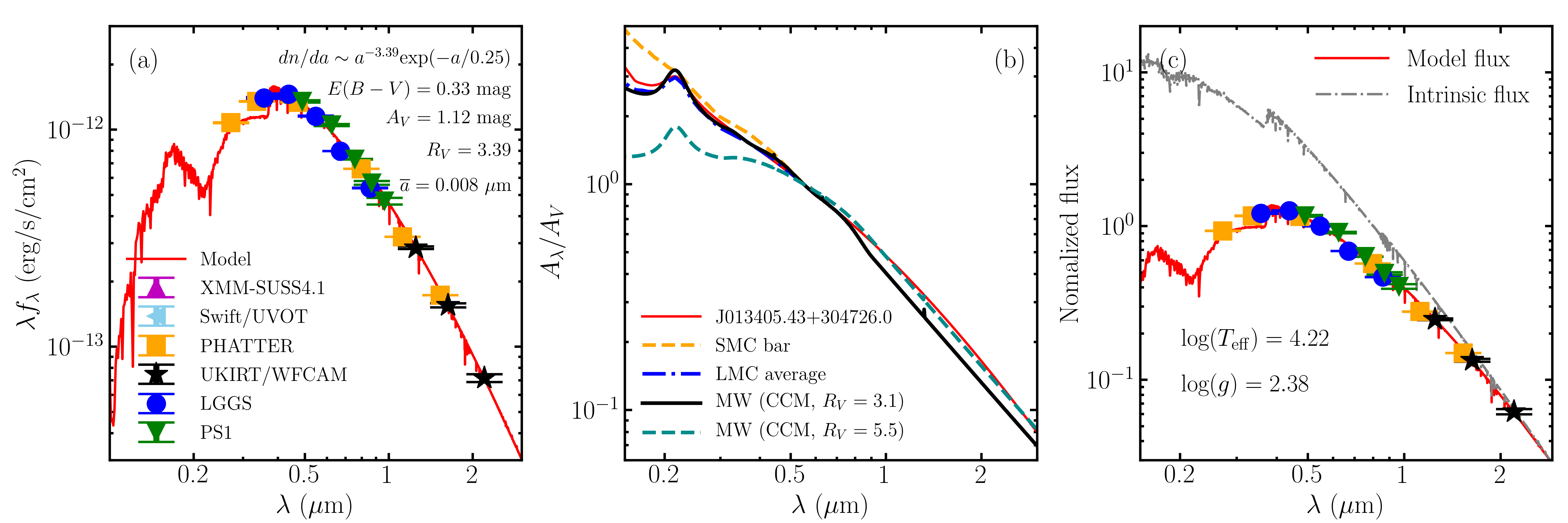}
	\caption{Example results for B-type supergiant J013405.43+304726.0.
	Panel(a) compares the model SED with the observed photometry.
	Panel(b) presents the extinction curve toward the sight line of this star compared with those of the average LMC, the bar region of SMC and the MW.
	Panel(c) shows the normalized intrinsic SED and the best-fitting SED compared to the observed data with the effective temperature log$(T_{\rm eff})$ and the surface gravity log$(g)$ marked.
	\label{fig:example}}
\end{figure}

With the model SEDs fit to the observed data, the fitting parameters for each tracer, including $\alpha$ in the dust model, extinction in the $V$ band $A_V$, the effective temperature log$(T_{\rm eff})$ and the surface gravity log$(g)$, can be derived, as well as the corresponding extinction curve, the average dust radius $\overline{a}$, the color excess $E(B-V)$, and the total-to-selective extinction ratio $R_V$.
The average dust radius $\overline{a}$ is derived based on equation (8) in \citet{2016PandSS..133...36N}.
$E(B-V)$ is derived by $E(B-V) = A_V(A_B/A_V-1)$, and then $R_V = A_V/E(B-V)$.
%\textbf{It is found that the variation in the mass ratio of graphite to silicate ($f_{cs}$) does not substantially change the derived extinction results, indicating that the results for M33 is not affected by the abundances of silicate and graphite in the dust. 
%In this work, we adopt $f_{cs} = 0.3$ for all the tracers, because the model SEDs derived from the dust model with $f_{cs} = 0.3$ generally fit the observed SED better.}
The results for one star in the extinction sample are plotted in Figure \ref{fig:example} as an example to show the comparison of the best-fitting model SED to the observed data with fitting parameters $\alpha$, $A_V$, the derived parameters $R_V$, $E(B-V)$ and $\overline{a}$ marked, the corresponding extinction curve and the comparison of normalized model SED and normalized intrinsic SED.

% ==========

\subsection{Results Screening} \label{subsec:result_select}

As mentioned in \citetalias{2022ApJS..259...12W}, \emph{pcFlag} is also introduced in this work to show the coverage of passbands adopted in the calculation for each tracers and the sample of the extinction tracers in this work is therefore divided into four subsamples (\emph{pcFlag} = `\emph{UVI}', \emph{pcFlag} = `\emph{UV}', \emph{pcFlag} = `\emph{VI}', \emph{pcFlag} = `\emph{V}')\footnote{\emph{U} in \emph{pcFlag} means the results are derived with UV data (here, this refers to the passbands bluer than the $U$ band), while \emph{V} and \emph{I} in \emph{pcFlag} are short for visual bands and near-IR bands (here, this refers to the passbands redder than the  $y$ band), respectively.}.
The numbers of the tracers in the four subsamples are 58, 20, 73 and 192, respectively.
It is considered that the results for the sight lines with \emph{pcFlag} = `\emph{UVI}' are the most reliable, which are adopted to analyze the extinction curves in this work.
The criteria used to select the reasonable results are similar to those in \citetalias{2022ApJS..259...12W}, as follows:
%As the results for all the tracers are presented, we need to eliminate unreliable and unreasonable tracers.
%The selection criteria are similar to those in \citetalias{2021arXiv211109523W}, as follows:

I. The acceptance fraction, which indicates the fraction of proposed steps that are accepted in the MCMC process, is in the range of 0.2 - 0.5 \citep{2013PASP..125..306F}.
%According to \citet{2013PASP..125..306F}, the autocorrelation time is the best indicator of MCMC performance, of which the acceptance fraction is the easiest and simplest proxies \citep{2013PASP..125..306F}.
%If things are going well, it should be in the range of 0.2-0.5.

II. The derived color excess $E(B-V)$ is larger than the foreground extinction for M33 [$E(B-V) > 0.06$ mag] because slightly reddened stars may lead to larger errors \citep{2015ApJ...815...14C}.

III. The derived total-to-selective extinction ratio is in the range of $R_V = 1.5 - 7$.
In our extinction sample, there are also some reddened stars with good MCMC performance, yet the derived values of $R_V$ are unusually as high ($R_V > 7$) as those in \citetalias{2022ApJS..259...12W}.
The results may lead to acceptable models that fit the observations well, but the derived values of $R_V$ are unphysical because the values of $R_V$ are observationally in the range of $ 2 \lesssim R_V \lesssim 6$ \citep{1990ARA&A..28...37M,1992ApJ...393..193W,1999PASP..111...63F,2011piim.book.....D,2017ApJ...848..106W}.

There are 58 tracers with \emph{pcFlag} = `\emph{UVI}' in the extinction sample, of which 39 tracers are selected for the further analysis based on the selection criteria mentioned above.
It is anticipated that sufficient data covering UV to near-IR bands will bring more reliable results, because UV and near-IR data can effectively constrain the extinction curve and the intrinsic spectra.
Although the results with \emph{pcFlag} = `\emph{UVI}' are recommended in this work, we also present the selected results for the tracers from the other three subsamples in the following section (see Table \ref{tab:re} and Table \ref{tab:re_dis}) as a reference.
%Based on the selection criteria mentioned above, 9 O-type and 117 B-type supergiants were selected from the sample with reliable results for further analysis.

%There are 217 removed sight lines, of which more than 80\% are due to the unphysical $R_V$.
%In addition, about 90\% of the sight line with unphysical $R_V$ lack UV data.
%

% ===========================================

\section{Results and discussion} \label{sec:re}

\subsection{The Extinction Curves in M33} \label{subsec:re}

The results for each selected tracer are partly listed in Table \ref{tab:re}, of which the entirety is available in machine-readable form.
The ID and spectral type for each tracer are extracted from the LGGS catalog \citep{2016AJ....152...62M}.
The fitting parameter [$\alpha$, log$(T_{\rm eff})$, log$(g)$ and $A_V$] and the uncertainties are derived based on the 50\%, 16\% and 84\% values of the parameter spaces generated from the EMCEE results.
The corresponding values of $\overline{a}$, $E(B-V)$ and $R_V$ derived in Section \ref{subsec:fitting} are also listed in Table \ref{tab:re}.
The columns named ``Total bands'' and ``pcFlag'' in Table \ref{tab:re} present the number of passbands and the coverage of the passbands adopted in the calculation.
The columns following the ``pcFlag'' column are the photometric information in each band, which are available in a machine-readable format.
It should be noted that although extinction curves with smaller $R_V$ values usually have stronger 2175 $\,{\rm \AA}$ bumps and steeper far-UV rises, the CCM extinction law with only one parameter $R_V$ is not generally applied in external galaxies \citep{2015ApJ...815...14C}; thus, $R_V$ cannot completely describe the extinction features on the extinction curves in external galaxies (\citetalias{2022ApJS..259...12W}).
As a result, we prefer to adopt the parameter $\alpha$ in the dust size distribution function to describe the dust extinction and dust properties in M33 in this work.

Table \ref{tab:re_dis} summarizes the median values with the upper and lower limits of each parameter.
As mentioned in Section \ref{subsec:result_select}, we divide our extinction sample into four subsamples based on the coverage of passbands adopted in the calculation (\emph{pcFlag}). Lines 1 to 4 in Table \ref{tab:re_dis} present the results of the four subsamples, and the fifth line shows the results of all the selected tracers in the extinction sample.
As illustrated in \citetalias{2022ApJS..259...12W}, the results for the sight lines with \emph{pcFlag = `UVI'} are the most reliable and are consequently adopted to describe the general extinction law in M33.
Lines 6 to 9 in Table \ref{tab:re_dis} show the influence of the lack of UV or near-IR data on the results (see Section \ref{subsec:UVIR} for details).
The last line in Table \ref{tab:re_dis} is the results of fitting our model extinction curves derived directly from the silicate-graphite dust model to the MW extinction curve \citep[\citetalias{2019ApJ...886..108F} hereafter]{2019ApJ...886..108F} for comparison.
%(\citealt{2019ApJ...886..108F}, F19 hereafter)

\begin{figure}
	\centering
	\includegraphics[scale=0.8]{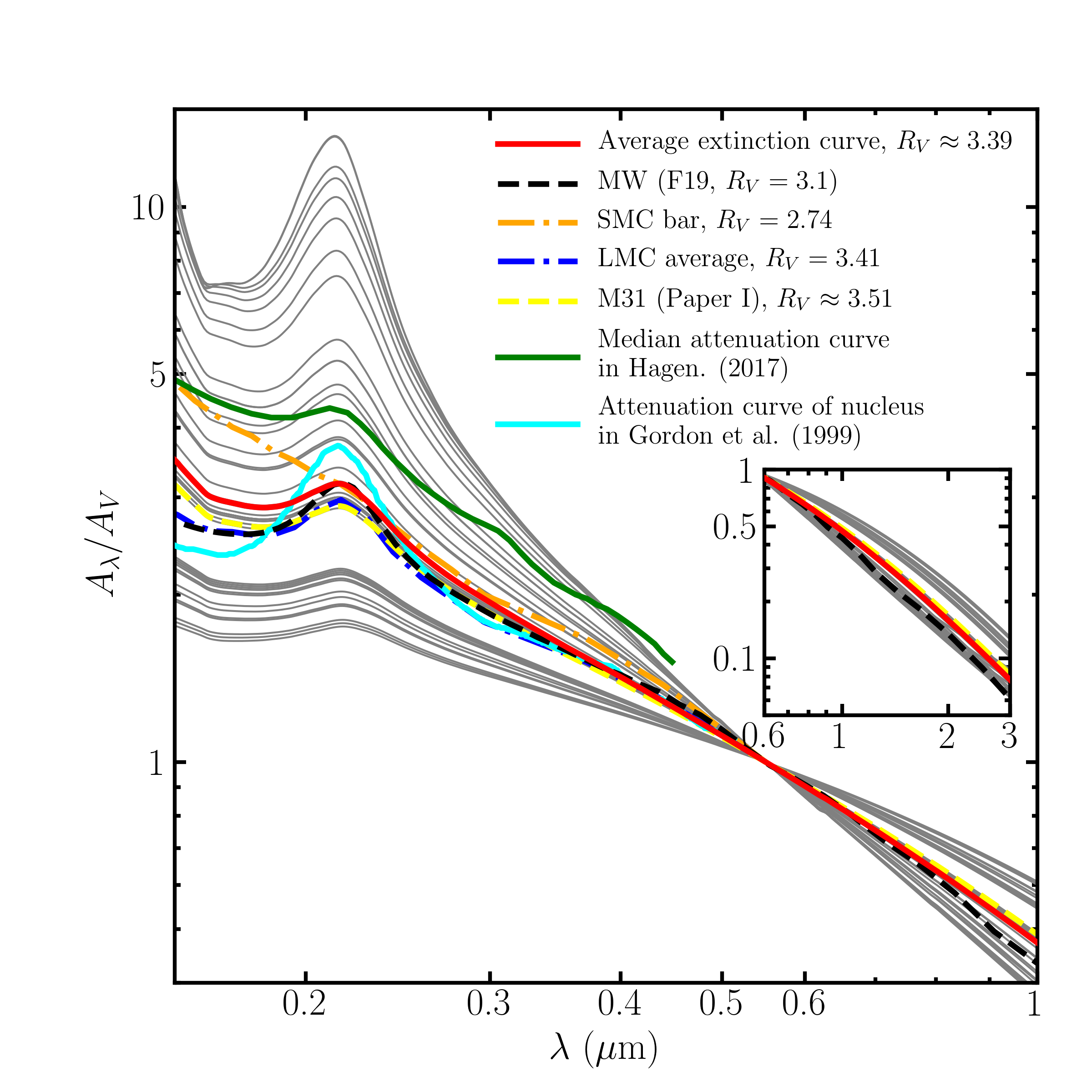}
	\caption{The dust extinction curves toward the sight lines with \emph{pcFlag = `UVI'} are plotted as solid gray lines.
	The main figure shows the extinction laws from UV (0.15 $\mu{\rm m}$) to 1 $\mu{\rm m}$, and the inset shows those from 0.6 $\mu{\rm m}$ to near-IR bands (3 $\mu{\rm m}$).
	The median extinction curve derived in this work (solid red line) is compared with the median attenuation curve derived in \citet{2017PhDT.......221H} (solid green line) and the attenuation curve of M33 nucleus derived in \citet{1999ApJ...519..165G} (solid cyan line).
	The black dashed line shows the extinction law toward the diffuse region in the MW.
	The yellow dashed line is the average extinction curve in M31 derived in \citetalias{2022ApJS..259...12W}.
	The orange dashed-and-dotted line and the blue dashed-and-dotted line are the extinction curve of the SMC bar and the average LMC extinction curve \citep{2003ApJ...594..279G}, respectively.
	\label{fig:M33ext}}
\end{figure}

The extinction curves toward the sight lines with \emph{pcFlag = `UVI'} are plotted in Figure \ref{fig:M33ext} with gray solid lines.
The extinction curves in M33 cover a wide range of shapes, from flat extinction curves with large $R_V$ to steep curves with obvious 2175 $\,{\rm \AA}$ bumps, indicating the inhomogeneous interstellar environment and dust distribution in M33.
The red solid line in Figure \ref{fig:M33ext} shows the average extinction derived based on the median value of the fitting parameter $\alpha$ for the sight lines with \emph{pcFlag = `UVI'}.
The average extinction curve in M33 shows similarity to the extinction curve in the diffuse region of the MW and the average LMC extinction curve, but with a slightly weaker 2175 $\,{\rm \AA}$ bump and a slightly steeper rise in the UV bands.
The average extinction law in M33 derived in this work can be applied to the general extinction correction in M33, and those toward individual sight lines can help with high-precision extinction correction (see Section \ref{subsec:prediction} for details).

\begin{figure}
	\centering
	\includegraphics[scale=0.8]{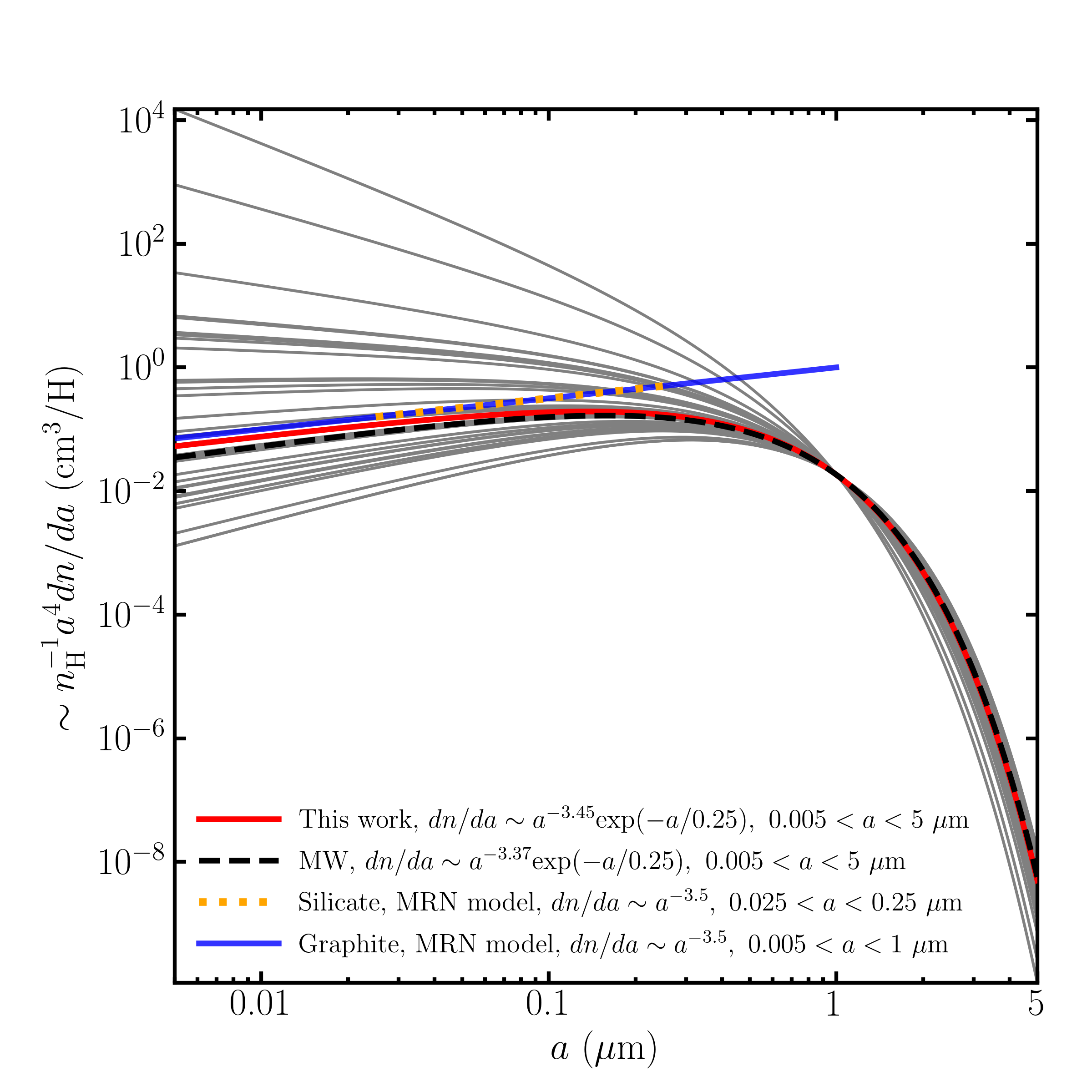}
	\caption{Dust size distributions for the selected tracers with \emph{pcFlag} = 'UVI'.
	The solid red line indicates the average dust size distribution for M33 derived in this work.
	The black dashed line shows the corresponding dust size distribution for the diffuse region in the MW as fitting the model extinction curves to the F19 extinction law ($R_V = 3.1$).
	The yellow dotted line and the solid blue line are the typical MRN dust size distributions for silicate and graphite, respectively \citep{1977ApJ...217..425M}.
	\label{fig:pm_dis}}
\end{figure}

The dust size distributions toward the selected tracers with \emph{pcFlag = `UVI'} are plotted in Figure \ref{fig:pm_dis} with gray solid lines.
As we know, grains with sizes that are comparable to the wavelength absorb and scatter light most effectively ($2\pi a/\lambda \approx 1$, where a is the spherical radius of the grain, \citealt{2009shm..book.....L}).
Interstellar dust has long been considered to be ``submicron-sized'' ($\approx 0.1~\mu {\rm m}$) since dust extinction was first confirmed by \citet{1930PASP...42..214T},
because grain models that reproduce the observed extinction should have extinction in the visible bands ($\lambda \approx 0.55~\mu {\rm m}$) dominated by grains with $a \approx 0.1~\mu {\rm m}$ \citep{2011piim.book.....D}.
However, it is now well recognized that the dust size actually spans a wide range from subnanometers to micrometers \citep{2015ApJ...811...38W}.
In addition, the strong rise to $\lambda \approx 0.1~\mu{\rm m}$ on the extinction curves requires a large abundance of grain with $2\pi a/\lambda \lesssim 1$; thus, interstellar dust must include a large population of grains with $a \lesssim 0.015~\mu{\rm m}$.
Based on equation (8) in \citet{2016PandSS..133...36N}, we derive that the average dust size $\overline{a}$ toward individual sight lines with \emph{pcFlag = `UVI'} is in the range of 5.78-9.64 nm.
The median value of the average dust size is $\overline{a} \approx 7.54$ nm, which is smaller than those of the MW ($\overline{a_{\rm MW}}~\approx~8.36$ nm) and M31 ($\overline{a_{\rm M31}}~\approx~8.41$ nm).

% ==========

%High effective temperatures as they are because of their early spectral types, the peak wavelengths of their blackbody-radiation-like intrinsic spectra should be around or shorter than UV bands.
%For those early type supergiants with high effective temperatures, the derived intrinsic spectra will be unreliable if the UV observed photometry centering at relatively shorter wavelengths is absent to restrict the shape of the whole SEDs and so do the derived extinction curves.

\movetabledown=1.5in
\begin{rotatetable}
	\begin{deluxetable}{ccccccccccccccccc}
		\tablecaption{An extracted list of the selected tracers in the sample with spectral type, fitting parameters ($\alpha,~{\rm log}~T_{\rm eff},~{\rm log}~g,~A_V$), derived $\overline{a}$, $E(B-V)$, $R_V$ and information about observed photometry adopted in this work$^a$.
		\label{tab:re}}
		\tablehead{
		\colhead{LGGS ID} & \colhead{LGGS SpT} & \colhead{$\alpha^{b,c}$} & \colhead{${\rm log}(T_{\rm eff})^c$} & \colhead{${\rm log}(g)^c$} & \colhead{$A_V$$^c$} &\colhead{$\overline{a}^d$} & \colhead{$E(B-V)$} & \colhead{$R_V$} &  \colhead{Total bands} & \colhead{pcFlag$^e$} & \colhead{...}	\\	
		\colhead{} & \colhead{} & \colhead{} & \colhead{} & \colhead{} & \colhead{(mag)}  & \colhead{(nm)} &  \colhead{(mag)} & \colhead{} &  \colhead{} & \colhead{}	& \colhead{ }
		}
		\startdata
		J013244.40+301547.7 & B6I & $3.46_{-0.04}^{+0.04}$ & $4.05_{-0.01}^{+0.01}$ & $2.26_{-0.09}^{+0.1}$ & $0.76_{-0.04}^{+0.04}$ & 7.53 & 0.23 & 3.24 & 11 & UVI \\
		J013250.65+304005.3 & B3I & $2.93_{-1.08}^{+0.05}$ & $4.2_{-0.02}^{+0.01}$ & $2.54_{-0.07}^{+0.06}$ & $0.37_{-0.05}^{+0.28}$ & 8.89 & 0.07 & 5.01 & 5 & V \\
		J013256.37+303552.1 & B1Ia & $3.04_{-0.31}^{+0.2}$ & $4.3_{-0.02}^{+0.02}$ & $2.74_{-0.07}^{+0.07}$ & $0.35_{-0.04}^{+0.03}$ & 8.51 & 0.08 & 4.53 & 16 & UVI \\
		J013256.61+302740.6 & B1.5Ia+Neb & $3.37_{-0.05}^{+0.09}$ & $4.3_{-0.01}^{+0.01}$ & $2.73_{-0.07}^{+0.07}$ & $0.35_{-0.03}^{+0.03}$ & 7.69 & 0.10 & 3.46 & 10 & V \\
		J013307.49+303042.7 & B7I & $4.35_{-0.20}^{+0.14}$ & $4.11_{-0.01}^{+0.02}$ & $2.26_{-0.09}^{+0.09}$ & $0.26_{-0.04}^{+0.04}$ & 6.46 & 0.13 & 1.93 & 16 & UVI \\
        \enddata
		\tablecomments{$^a$ This is an extracted table, and the entire table is available in machine-readable form.\\
		$^b$ The dust size distribution can be expressed as follows:
		$dn/da \sim a^{-\alpha}{\rm exp}(-a/0.25),~0.005 < a < 5 \mu {\rm m}$, as described in Section \ref{sec:method}.\\
		$^c$ The final results and the uncertainties for each parameter are derived from the 50th, 16th and 84th percentiles of the parameter spaces from the EMCEE results.\\
		$^d$ The average dust radius $\overline{a}$ is calculated from equation (8) in \citet{2016PandSS..133...36N}.\\
		$^e$ This is the flag for the coverage of passbands adopted in the calculation.
		U = UV bands (here, this refers to the passbands bluer than the $U$ band).
		V = Visual band (here, this refers to the passbands from the $U$ to $y$ bands).
		I = IR band (here, this refers to the passbands redder than the $y$ band).
		}
	\end{deluxetable}
\end{rotatetable}

% ==

\begin{deluxetable*}{cccccccccc}
	\tablecaption{General results of $\alpha$, $A_V$, $\overline{a}$, $E(B-V)$ and $R_V$ with the corresponding upper and lower limits for M33$^a$. \label{tab:re_dis}}
		\tablehead{	
		\colhead{} & \colhead{Total Sample} &  \colhead{$\alpha$} & \colhead{$A_V$} & \colhead{$\overline{a}$} & \colhead{$E(B-V)$} & \colhead{$R_V$} \\
		\colhead{} & \colhead{} &  \colhead{} & \colhead{(mag)} & \colhead{(nm)} & \colhead{(mag)} & \colhead{}
		}
	\startdata
	    pcFlag = `UVI'$^b$ & 39 & $3.45_{-0.70}^{+2.37}$ & $0.43_{-0.28}^{+1.40}$ & $7.54_{-1.76}^{+2.10}$ & $0.13_{-0.06}^{+0.29}$ & $3.39_{-1.51}^{+2.92}$ \\
		pcFlag = `UV' &	14 & $3.36_{-0.59}^{+2.11}$ & $0.54_{-0.30}^{+1.04}$ & $7.71_{-1.83}^{+1.85}$ & $0.15_{-0.07}^{+0.32}$ & $3.39_{-1.44}^{+2.30}$ \\
		pcFlag = `VI' & 32 & $3.39_{-0.65}^{+1.71}$ & $0.45_{-0.32}^{+0.51}$ & $7.65_{-1.62}^{+2.06}$ & $0.13_{-0.07}^{+0.07}$ & $3.39_{-1.52}^{+2.92}$ \\
		pcFlag = `V' & 41 & $3.31_{-0.64}^{+2.31}$ & $0.47_{-0.33}^{+1.44}$ & $7.82_{-1.99}^{+2.24}$ & $0.13_{-0.07}^{+0.28}$ & $3.65_{-1.78}^{+2.65}$ \\
		All tracers & 126 & $3.38_{-0.72}^{+2.43}$ & $0.45_{-0.32}^{+1.45}$ & $7.66_{-1.88}^{+2.41}$ & $0.13_{-0.07}^{+0.33}$ & $3.39_{-1.52}^{+2.92}$ \\
		\hline
		Tracers with near-IR data$^c$ &	71 & $3.45_{-0.71}^{+2.37}$ & $0.45_{-0.32}^{+1.38}$ & $7.54_{-1.76}^{+2.17}$ & $0.13_{-0.07}^{+0.29}$ & $3.39_{-1.52}^{+2.92}$ \\
		& 51$^d$ & $3.45_{-0.79}^{+2.00}$$^{\ast}$ & $0.55_{-0.40}^{+1.10}$$^{\ast}$ & $7.53_{-1.64}^{+2.57}$$^{\ast}$ & $0.14_{-0.07}^{+0.29}$$^{\ast}$ & $3.26_{-1.36}^{+3.32}$$^{\ast}$ \\
		Tracers with UV data$^c$ & 53 & $3.45_{-0.70}^{+2.37}$ & $0.45_{-0.30}^{+1.38}$ & $7.54_{-1.76}^{+2.10}$ & $0.13_{-0.06}^{+0.33}$ & $3.39_{-1.51}^{+2.92}$ \\
		& 36$^d$ & $3.24_{-0.49}^{+1.89}$$^{\ast}$ & $0.46_{-0.22}^{+1.10}$$^{\ast}$ & $7.98_{-1.96}^{+1.65}$$^{\ast}$ & $0.15_{-0.09}^{+0.26}$$^{\ast}$ & $3.84_{-1.94}^{+2.14}$$^{\ast}$ \\
		\hline
		MW (F19, $R_V = 3.1$)$^e$ & & 3.37 & & 8.36
	\enddata
	\tablecomments{$^a$ The superscript and the subscript in the table indicate the derived upper limit value and lower value of the derived parameters extracted from Table \ref{tab:re} for the selected tracers.\\
	$^b$ The results of the sight lines with \emph{pcFlag = `UVI'} are adopted to derive the general extinction law in M33.\\
	$^c$ For tracers with near-IR (UV) data, the calculation is repeated without taking near-IR (UV) data into consideration, and reliable results are listed with $^{\ast}$ for comparison.\\
	$^d$ For the 71 (53) tracers with near-IR (UV) data, when repeating the calculation without adopting near-IR (UV) data, reliable results can be derived for only 51 (36) tracers.\\
	$^e$ The Levenberg-Marquardt method is adopted to fit the model extinction curves to the F19 extinction curve with $R_V = 3.1$.
	There is only one parameter ($\alpha$) in our model extinction curves, and a grid ranging from 0.50 to 7.00 with a step of 0.01 is taken.\\
	}
\end{deluxetable*}

% ==

% ==

% ==========

\subsection{Comparison with the Local Galaxies}

We compare the average extinction curve in M33 with those in the MW and other local group galaxies (SMC, LMC, M31) in Figure \ref{fig:M33ext}.
The average LMC extinction law \citep{1978Natur.276..478N,1986AJ.....92.1068F,2003ApJ...594..279G} resembles that in the MW, while most of the extinction curves in the SMC bar region \citep{1984A&A...132..389P,2003ApJ...594..279G} display a nearly linear rise with $\lambda^{-1}$ and an absent 2175 $\,{\rm \AA}$, similar to those in the starburst galaxies \citep{1994ApJ...429..582C}.
The average extinction curve in M31 derived in \citetalias{2022ApJS..259...12W} (yellow dashed lines in Figure \ref{fig:M33ext}) shows similarity to that of the MW but rises less steeply in the far-UV bands.
The average extinction curve in M33 is similar to that of M31 in shape but with a slightly larger slope.

The average dust extinction curve derived in this work is also compared with the attenuation curves derived in \citet{1999ApJ...519..165G} and \citet{2017PhDT.......221H} in Figure \ref{fig:M33ext}.
As illustrated in Section \ref{sec:intro}, attenuation curves include both extinction and the assumed geometry of dust and stars, so attenuation is aimed at the effect of dust on an area instead of an individual sight line.
\citet{1999ApJ...519..165G} adopted radiative transfer modeling from UV to NIR of the M33 nucleus, which is an ideal interstellar environment of starburst, and found an MW-like attenuation curve with a strong 2175 $\,{\rm \AA}$ bump (solid cyan line in Figure \ref{fig:M33ext}).
\citet{2017PhDT.......221H} modeled the SEDs for 1170 large pixels in M33 from FUV to NIR and derived a steep median attenuation curve with a weaker 2175 $\,{\rm \AA}$ bump (solid green line in Figure \ref{fig:M33ext}).

\begin{figure}[ht!]
	\includegraphics[scale=1]{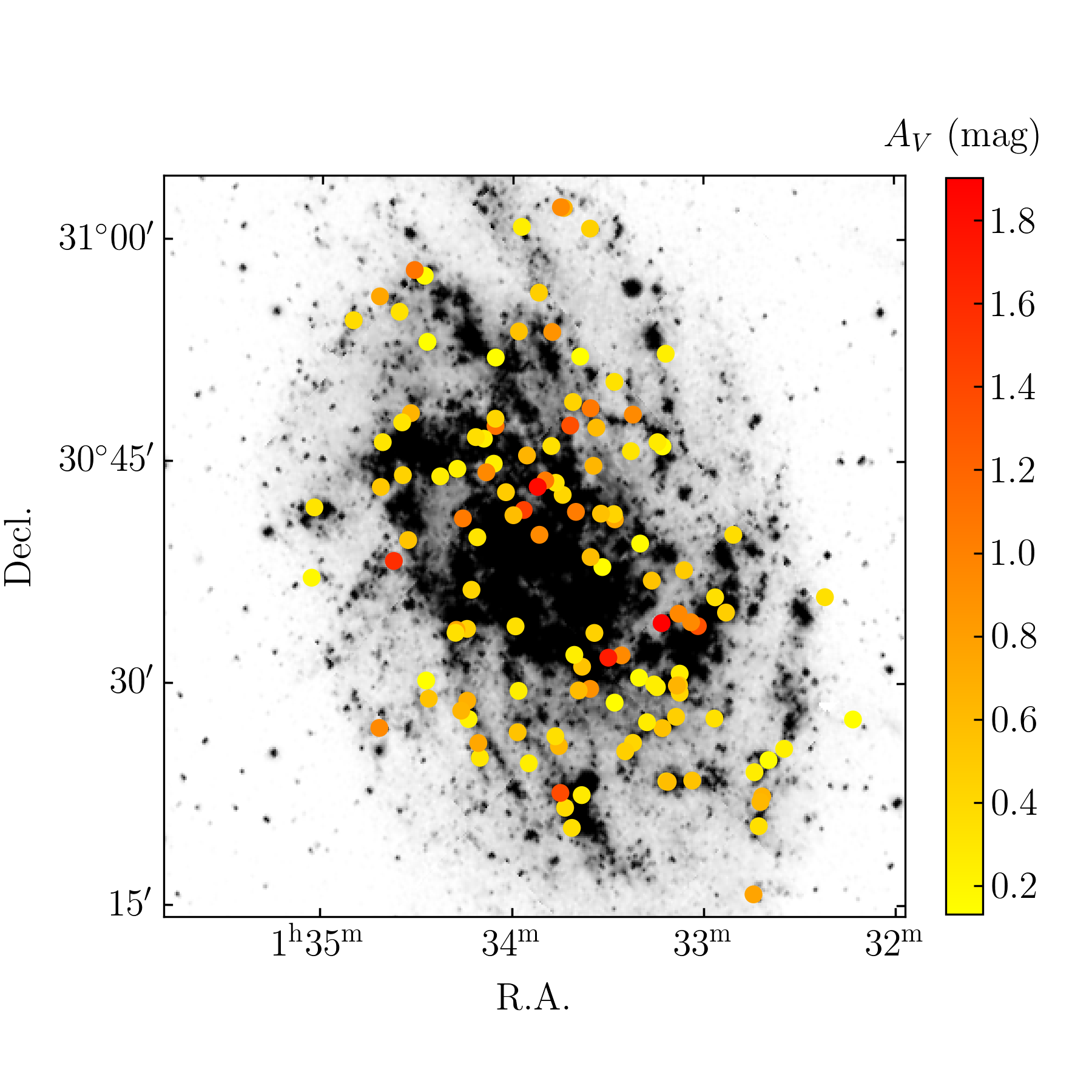}
	\caption{The $A_V$ distribution derived in this work.
	The background image is taken from the GALEX ultraviolet observation.
	 \label{fig:av_dis} }
\end{figure}

The average extinction curve in M33 derived in this work presents a similar slope to \citet{1999ApJ...519..165G} but with a weaker 2175 $\,{\rm \AA}$ bump as the median one in \citet{2017PhDT.......221H}.
In this work, we map the derived $A_V$ of the selected tracers in Figure \ref{fig:av_dis} and find that the median value of $A_V$ is $\approx$ 0.43 mag, which is slightly smaller than the median $A_V$ ($\approx 0.53$ mag) derived in \citep{2017PhDT.......221H} and larger than the mean amount of dust extinction ($A_V \approx 0.25$ mag) measured in \citet{2009A&A...493..453V}.
The discrepancy may be due to the different scales of dust and the different stellar models \citep{2013ARA&A..51..393C}.
In addition, we eliminate the results with $E(B-V) < 0.06$ mag, as mentioned in Section \ref{subsec:result_select}, because slightly reddened stars may lead to larger errors.
As a result, tracers with small $A_V$ values are excluded, increasing the median value of $A_V$.

\citet{2022AJ....163...16M} combined the Starburst99 \citep{1999ApJS..123....3L}+YGGDRASIL \citep{2011JMatR..26.1260Z} simple stellar population models and the starburst attenuation curve \citep{2000ApJ...533..682C} to model the SEDs of the young star cluster population in M33.
They found that all the star clusters have moderate-to-small internal extinction, i.e., all the star clusters have $E(B-V) < 0.6$ mag and approximately 2/3 of them have $E(B-V) < 0.2$ mag.
We also derive small extinction values for all the supergiants in our extinction sample, i.e., all the supergiants have $E(B-V) < 0.45$ mag, and approximately 2/3 of them have $E(B-V) < 0.2$ mag, which is consistent with \citet{2022AJ....163...16M}.

% ==========
\subsection{The 2175 $\,{\rm \AA}$ bump} \label{subsec:2175}
The 2175 $\,{\rm \AA}$ bump, which is the broad excess in the extinction curve at a rest wavelength $\lambda \approx 2175$ $\,{\rm \AA}$, is the strongest signature of dust in the interstellar medium \citep{2021ApJ...909..213K}.
It has been considered as a unique probe of the nature of dust in galaxies since it was discovered by \citet{1965ApJ...142.1683S}.
The 2175 $\,{\rm \AA}$ bump is obvious in the extinction curves toward the individual sight lines of the MW (e.g., \citealt{1986ApJ...307..286F,1990ApJS...72..163F}; \citetalias{2019ApJ...886..108F}), LMC \citep{1978Natur.276..478N,1986AJ.....92.1068F,2003ApJ...594..279G} and M31 (\citealt{2014ApJ...785..136D,2015ApJ...815...14C}; \citetalias{2022ApJS..259...12W}), while it was almost absent in the SMC \citep{1984A&A...132..389P,2003ApJ...594..279G}.
On galaxy scales, it was found that there is no significant 2175 $\,{\rm \AA}$ bump in the attenuation curves of nearby starburst galaxies \citep{1994ApJ...429..582C,1997ApJ...487..625G,2000ApJ...533..682C} and Lyman break galaxies at high redshifts ($z > 2$, \citealt{2003ApJ...587..533V}).
As a result, the attenuation curves with no 2175 $\,{\rm \AA}$ bump in \citet{2000ApJ...533..682C} are commonly adopted for both local and distant star-forming galaxies.
However, the 2175 $\,{\rm \AA}$ bump has been detected and even measured for star-forming galaixes by many recent works (e.g. \citealt{2007A&A...472..455N,2009A&A...499...69N,2011A&A...533A..93B,2012A&A...545A.141B,2015ApJ...800..108S,2017ApJ...851...90B,2018ApJ...859...11S,2020ApJ...888..108B,2020ApJ...899..117S}).

As to M33, although it was one of the star-forming galaxies in \citet{1994ApJ...429..582C} with no significant 2175 $\,{\rm \AA}$ bump in the attenuation curve, recent studies \citep{1999ApJ...519..165G,2017PhDT.......221H} indicated that there exists a 2175 $\,{\rm \AA}$ bump in the attenuation curve.
Since graphite is one of the possible candidates of the carriers of the 2175 $\,{\rm \AA}$ bump \citep{1965ApJ...142.1681S}, we adopt the silicate-graphite dust model in this work to derive the overall extinction curves from UV to near-IR toward individual sight lines in M33.
%In this work, the 2175 $\,{\rm \AA}$ bump can be only modeled with the photometric points in $UVW2$, $UVM2$, $FUV$ and $NUV$ bands, which cannot be obtained for most of the tracers in the extinction sample because of the observation limitation in the UV bands.
%As a result, it is difficult to analyze the UV extinction curves in detail (e.g., the strength and the FWHM of the 2175 $\,{\rm \AA}$ bump), and we just adopt the silicate-graphite dust model with the fixed mass ratio of graphite to silicate $f_{cs}$ to derive the overall extinction curves from UV to near-IR toward individual sight lines in M33.
In order to find out whether the 2175 $\,{\rm \AA}$ bump really exists in the extinction curves of M33, we also adopt the model extinction curves without a 2175 $\,{\rm \AA}$ bump
derived from the silicate dust model ($f_{cs} = 0$, no carbonaceous grains) to repeat the calculation for the tracers with ultraviolet data.
%As graphite is one of the possible candidates of the carriers of the 2175 $\,{\rm \AA}$ bump \citep{1965ApJ...142.1681S}, the model extinction curves without a 2175 $\,{\rm \AA}$ bump are derived from the silicate dust model ($f_{cs} = 0$, no carbonaceous grains).
By comparing the median values of $\chi^2/d.o.f.$\footnote{$\chi^2/d.o.f. = \frac{1}{N_{\rm data}-N_{\rm para}}\sum{\frac{[{\rm log}(f_{\rm model})-{\rm log}(f_{\rm observed})]^2}{\sigma^2}}$, where $N_{\rm data}$ is the number of the observed photometric points adopted in the calculation, $N_{\rm para}$ is the number of adjustable parameters (see Section \ref{subsec:model_sed} for details), $f_{\rm observed}$ is the observed flux of the photometric point, $f_{\rm model}$ is the model flux of the photometric point and $\sigma$ is the difference between logarithm of extreme and logarithm of $f_{\rm observed}$.} derived from both dust models for each tracer, it is found that the extinction curves derived from the silicate-graphite dust model can generally recover the observed SEDs better than the silicate dust model.
%it is inferred that for most of the tracers with the UV data,
We therefore suggest that there exists a 2175 $\,{\rm \AA}$ bump in the extinction curves of M33.
%we recommend the extinction curves with a 2175 $\,{\rm \AA}$ bump for M33 in this work.
The fine structure of the UV extinction curves can be analyzed and more comprehensive results can be expected, if the UV data is adequate in the future.

% ==========

\subsection{Influence of IR and UV Photometry} \label{subsec:UVIR}

Photometric data that cover a wider range of passbands will constrain the observed SED better and bring more reliable results.
Because of the observation limit, a number of tracers lack photometry in the $UVW2, UVM2, UVW1$ and PHATTER bands.
Meanwhile, UKIRT data are also not applied to all tracers in the extinction sample, as mentioned in Section \ref{sec:data}.
It is thus necessary to determine whether the lack of photometry in the UV and near-IR bands affects the derived extinction law.

Lines 6 and 8 in Table \ref{tab:re_dis} summarize the results of the selected tracers with near-IR data and with UV data, respectively.
We repeat the calculation for these two groups of tracers but ignore the near-IR data or UV data and list the number of tracers with reliable results as well as the derived results in the seventh and ninth lines of Table \ref{tab:re_dis}, respectively.
As shown in Table \ref{tab:re_dis}, the number of tracers with reliable results is significantly reduced when UV data or near-IR data are not adopted in the calculation, indicating that photometric data in wider bands bring more reliable results.

\begin{figure}[ht!]
	\includegraphics[scale=0.5]{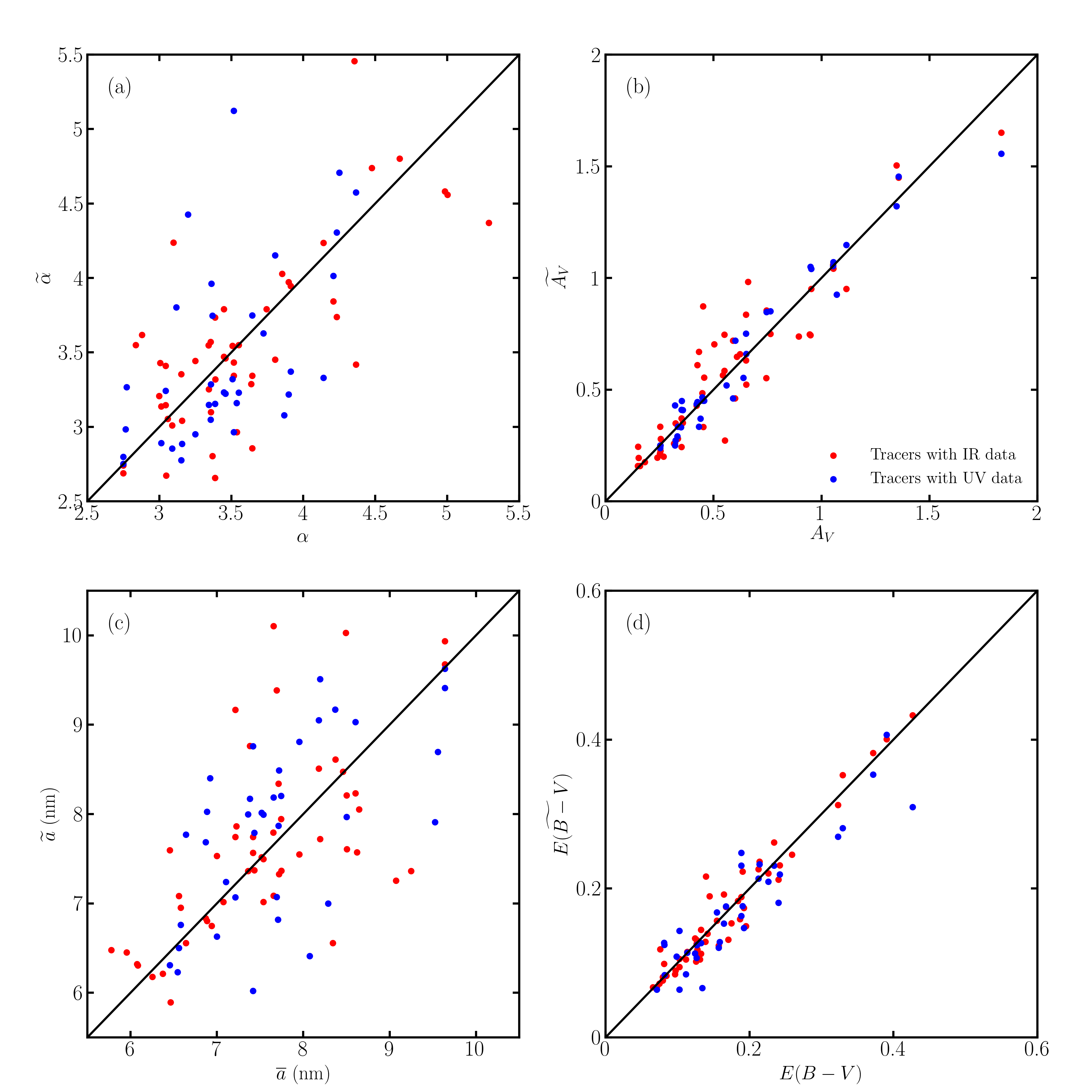}
	\caption{Influence of lacking UV or near-IR data on the derived $\alpha,~A_V$, $\overline{a}$ and $E(B-V)$.
	The red and blue dots indicate the tracers in the seventh line and the ninth line in Table \ref{tab:re_dis}, respectively.
	The x-axis values for a certain red (blue) dot are the derived parameters [$\alpha,~A_V,~\overline{a},~E(B-V)$] with all available photometric data considered, while the corresponding y-axis values are the derived parameters [$\widetilde{\alpha},~\widetilde{A_V},~\widetilde{a},~\widetilde{E(B-V)}$] when repeating the calculation process but without taking the available near-IR (UV) data into consideration.
	The deviation between the red (blue) dots and $y=x$ presents the influence of the lack of near-IR (UV) data.
	 \label{fig:UVIR} }
\end{figure}

On the other hand, we compare the reliable results derived with near-IR (UV) data ignored for the tracers with near-IR (UV) data [Tracers in Line 6 (8) of Table \ref{tab:re_dis}] and the results extracted from Table \ref{tab:re} for the same tracers in Figure \ref{fig:UVIR}.
As Figure \ref{fig:UVIR} shows, the lack of UV or near-IR data has little impact on $A_V$ and $E(B-V)$ in this work.
However, the dust size parameter $\alpha$ and the average dust size $\overline{a}$ for most of the individual tracers are influenced by the coverage of the adopted passbands, indicating that UV and near-IR data are important to constrain the dust model.

To illustrate the reliability of the derived results for individual sight lines, as mentioned in Section \ref{sec:data}, we introduce \emph{pcFlag} in Table \ref{tab:re} to show the coverage of passbands adopted in the calculation.
The results of sight lines with \emph{pcFlag = `UVI'} are the most reliable, while those with \emph{pcFlag = `V'} are the least reliable.
We anticipate that the results could be more comprehensive if the observed data in multiple bands are adequate, especially in UV bands, because UV data can provide a strong constraint on the extinction model.
The coming 2 m-aperture Survey Space Telescope (also known as the China Space Station Telescope, CSST) will image approximately 17500 square degrees of the sky in the $NUV$, $u$, $g$, $r$, $i$, $z$ and $y$ bands \citep{2021CSB...111.111C} and will provide us with abundant data to explore the dust extinction law in M33 and other nearby star-resolved galaxies.

% ==========

\subsection{Prediction of Multiband Extinction} \label{subsec:prediction}

As mentioned in Section \ref{subsec:re}, the dust extinction curves derived in this work can help with the extinction correction in M33.
Based on the average extinction curve of M33 in this work, multiband extinction values from UV to near-IR are predicted, which are shown in Table \ref{tab:pre}.
High-precision extinction correction should refer to the extinction law of individual tracers derived in this work.
Table \ref{tab:pre_indi} partially lists the extinction values in multiple bands from UV to near-IR toward individual sight lines, and the entire table is available in machine-readable form.
The ID, spectral type, right ascension and declination listed in the first four columns of Table \ref{tab:pre_indi} for each tracer are obtained from the LGGS catalog \citep{2016AJ....152...62M}.
The $\alpha$, $\overline{a}$ and $E(B-V)$ in Table \ref{tab:pre_indi} are extracted from Table \ref{tab:re}.
The column named \emph{pcFlag} presents the passband coverage, as shown in Table \ref{tab:re}.
The following columns show the extinction values in multiple bands for each tracer.
Although the results for some tracers are not affected by the lack of UV or near-IR data, we recommend multiband extinction toward individual sight lines with \emph{pcFlag = `UVI'}.

When applying the method and results in this work, there are three aspects that need to be noted.
First, the extinction curves derived in this work are applicable from UV to near-IR ($\approx 3~\mu{\rm m}$) bands.
All dust models for the diffuse ISM predict that an extinction curve steeply declines with $\lambda$ at $1~\mu{\rm m} < \lambda < 7~\mu{\rm m}$ and increases at $\lambda > 7~\mu{\rm m}$ because of the $9.7~\mu{\rm m}$ silicate absorption feature \citep{1977ApJ...217..425M,1994ApJ...422..164K,2001ApJ...548..296W,2015llg..book...85L}.
However, many recent observations suggest that the extinction law in the mid-IR band ($3~\mu{\rm m} < \lambda < 8~\mu{\rm m}$) appears to be universally flat or gray in various interstellar environments \citep{1996A&A...315L.269L,1999ESASP.427..623L,2005ApJ...619..931I,2007ApJ...663.1069F,2009ApJ...707...89G,2009ApJ...696.1407N,2011ApJ...737...73F,2013ApJ...773...30W}.
Although the $\mu{\rm m}$-sized grain can be adopted to model the flat mid-IR extinction curve \citep{2015ApJ...811...38W}, it will make the model more complex and, thus, reduce the universality of the method.
As a result, the derived extinction curves cannot be applied to mid-IR bands at present.

Moreover, the classic silicate-graphite dust model adopted in this work may not be applied to analyze the fine structure of the extinction curves in UV bands, although it can be adopted to derive the overall reliable extinction curves in M33 from UV to near-IR bands.
As illustrated in Section \ref{subsec:2175}, the 2175 $\,{\rm \AA}$ bump is known to be an important feature of the extinction curves in UV bands, which was first discovered by \citet{1965ApJ...142.1683S}.
Since \citet{1965ApJ...142.1681S} pointed out that small graphite particles would produce absorption very similar to this observed feature, some form of graphitic carbon has been an attractive candidate because the $\pi \rightarrow \pi^{\ast}$ transition in graphite is responsible for the absorption feature at $\sim 2175 \,{\rm \AA}$ \citep{2003ARA&A..41..241D}.
However, this graphite hypothesis does not appear to explain the fact that the full width at half maxima (FWHM) of 2175 $\,{\rm \AA}$ varies with the interstellar environment while holding the central wavelength $\lambda_0$ nearly constant \citep{1993ApJ...414..632D}.
Currently, a polycyclic aromatic hydrocarbon (PAH) mixtures are carrier candidates for the 2175 $\,{\rm \AA}$ bump \citep{1992ApJ...393L..79J,2001ApJ...554..778L,2011ApJ...733...91X,2011ApJ...742....2S,2015ApJ...809..120M,2017ApJ...850..138M} because PAH molecules generally have strong $\pi \rightarrow \pi^{\ast}$ absorption in the 2000 - 2500 $\,{\rm \AA}$ region with variation in FWHM and small variation in $\lambda_0$.
As a result, we consider adding PAHs to the dust model in future work to analyze the fine structure of UV extinction curves and obtain a more detailed understanding of the dust properties in M33 and other nearby galaxies.

Finally, as shown in Figure \ref{fig:av_dis}, the size of the extinction sample adopted in this work is not adequate to cover the entire region of M33.
Thus, it can only provide us with a low-resolution extinction map to help with a rough extinction correction for certain regions in M33.
A major science project named CSST mentioned in Section \ref{subsec:UVIR} will provide us with larger extinction samples and adequate data in multiple bands.
We can expect further exploration of the dust properties and extinction law in M33 and in other nearby star-resolved galaxies, as well as the development of higher-precision extinction corrections in the near future.

\newpage

\startlongtable
\begin{deluxetable}{cccccccccc}
	\tablecaption{General extinction prediction for M33 in multiple bands from UV to IR. \label{tab:pre}}
		\tablehead{	
		\colhead{Band} & \colhead{$\lambda_{\rm eff}$$^a$} & \colhead{$A_{\lambda}/A_V$} & \colhead{$A_{\lambda}$$^b$ } \\
		\colhead{} & \colhead{$(\mu {\rm m})$} & & \colhead{(mag)}
		}
	\startdata	                
	$UVW2$/UVOT & 0.209 & $3.13$ & $1.4$ \\
	$UVM2$/UVOT & 0.225 & $3.04$ & $1.36$ \\
	$UVW1$/UVOT & 0.268 & $2.23$ & $1.0$ \\
	$F275W$/PHAT & 0.272 & $2.18$ & $0.98$ \\
	$NUV$/CSST & 0.29 & $2.02$ & $0.9$ \\
	$F336W$/PHAT & 0.336 & $1.71$ & $0.77$ \\
	$U$ & 0.357 & $1.6$ & $0.72$ \\
	$B$ & 0.443 & $1.27$ & $0.57$ \\
	$F475W$/PHAT & 0.473 & $1.19$ & $0.53$ \\
	$g$/CSST & 0.475 & $1.18$ & $0.53$ \\
	$V$ & 0.554 & $0.99$ & $0.45$ \\
	$r$/CSST & 0.612 & $0.88$ & $0.4$ \\
	$R$ & 0.67 & $0.79$ & $0.36$ \\
	$i$/CSST & 0.758 & $0.68$ & $0.31$ \\
	$F814W$/PHAT & 0.798 & $0.64$ & $0.29$ \\
	$I$ & 0.857 & $0.58$ & $0.26$ \\
	$z$/CSST & 0.911 & $0.54$ & $0.24$ \\
	$y$/CSST & 0.989 & $0.48$ & $0.22$ \\
	$F110W$/PHAT & 1.12 & $0.4$ & $0.18$ \\
	$J$/2MASS & 1.235 & $0.35$ & $0.16$ \\
	$F160W$/PHAT & 1.528 & $0.25$ & $0.11$ \\
	$H$/2MASS & 1.662 & $0.22$ & $0.1$ \\
	$K$/2MASS & 2.159 & $0.14$ & $0.06$ \\
	%$W1$/WISE & 3.353 & $0.06$ & $0.03$ \\
	%$[3.6]$/IRAC & 3.508 & $0.06$ & $0.03$ \\
	%$[4.5]$/IRAC & 4.437 & $0.04$ & $0.02$ \\
	%$W2$/WISE & 4.603 & $0.03$ & $0.02$
\enddata
	\tablecomments{$^a$ For the effective wavelengths of multiple bands (except CSST bands) used in this work refer to the SVO Filter Profile Service (http://svo2.cab.inta-csic.es/theory/fps/, \citealt{2012ivoa.rept.1015R}).
	The effective wavelengths of the CSST bands are calculated by $\lambda_{\rm eff} = \frac{\int \lambda^2T(\lambda)Vg(\lambda)d\lambda}{\int \lambda T(\lambda) Vg(\lambda)d \lambda}$, where $T(\lambda)$ is the filter transmission function and $Vg(\lambda)$ is the Vega spectrum.\\
	$^b$ $A_{\lambda}$ is the average extinction in M33 based on the median value of $A_V$.}
	\end{deluxetable}

\newpage

\movetabledown = 2in
\begin{rotatetable}
	\begin{deluxetable}{ccccccccccccccccc}
		\tablecaption{An extracted list of multiband extinction values for individual tracers$^a$. \label{tab:pre_indi}}
		\tablehead{	
		\colhead{LGGS ID} & \colhead{LGGS SpT} & \colhead{RA}  & \colhead{Dec} & \colhead{$\alpha$} & \colhead{$\overline{a}$} & \colhead{$E(B-V)$} & \colhead{pcFlag$^b$} & \colhead{$A_V$} & \colhead{$A_{UVW2}$}  & \colhead{$A_{UVM2}$} & \colhead{$A_{UVW1}$} & \colhead{$A_{F275W}$} &  \colhead{...}\\
		\colhead{} & \colhead{} & \colhead{(h:m:s)} & \colhead{(d:m:s)} & \colhead{} & \colhead{(nm)} &\colhead{(mag)} & \colhead{} & \colhead{(mag)} & \colhead{(mag)} & \colhead{(mag)} & \colhead{(mag)} & \colhead{(mag)}
		} 
\startdata
J013244.40+301547.7 & B6I & 01 32 44.37 & +30 15 47.6 & 3.46 & 7.53 & 0.23 & UVI & 0.76 & 8.37 & 7.3 & 3.23 & 3.08  \\
J013250.65+304005.3 & B3I & 01 32 50.62 & +30 40 05.2 & 2.93 & 8.89 & 0.07 & V & 0.37 & 4.1 & 3.58 & 1.58 & 1.51 \\
J013256.37+303552.1 & B1Ia & 01 32 56.34 & +30 35 52.0 & 3.04 & 8.51 & 0.08 & UVI & 0.35 & 3.9 & 3.41 & 1.51 & 1.44  \\
J013256.61+302740.6 & B1.5Ia+Neb & 01 32 56.58 & +30 27 40.5 & 3.37 & 7.69 & 0.1 & V & 0.35 & 3.87 & 3.38 & 1.49 & 1.43  \\
J013307.49+303042.7 & B7I & 01 33 07.46 & +30 30 42.6 & 4.35 & 6.46 & 0.13 & UVI & 0.26 & 1.2 & 1.14 & 0.72 & 0.7 \\
\enddata		
\tablecomments{$^a$ This is an extracted table for the same tracers as listed in Table \ref{tab:re} with a portion of the multiband extinction values.
The entire table is available in machine-readable form.\\
$^b$ Flag for the coverage of passbands adopted in the calculation (the same as Table \ref{tab:re})
}
\end{deluxetable}
\end{rotatetable}

% ====================
\newpage
% ====================

\subsection{Application in Other Nearby Star-Resolved Galaxies}

The method adopted in this work and \citetalias{2022ApJS..259...12W} is extended to other nearby star-resolved galaxies.
The LGGS provides $UBVRI$ plus the interference-image photometry of luminous stars in seven systems currently forming massive stars (IC 10, NGC 6822, WLM, Sextans A and B, Pegasus and Phoenix, \citealt{2007AJ....133.2393M}) in addition to the spiral galaxies M31 and M33 \citep{2006AJ....131.2478M,2016AJ....152...62M}.
We can isolate O-type and B-type supergiants in NGC 6822 and WLM from the LGGS catalog \citep{2007AJ....133.2393M}, which are selected as the extinction tracers.

The observed data for the tracers in NGC 6822 and WLM are from the LGGS catalog \citep{2007AJ....133.2393M}, the PS1 survey \citep{2016arXiv161205560C}, the UKIRT \citep{2013ASSP...37..229I} and the HST (\citealt{2017ApJS..230...24B}, only for NGC 6822).
We adopt $A_{V_0} = 0.646$ mag and $A_{V_0} = 0.104$ mag to remove the foreground extinction for NGC 6822 and WLM \citep{2011ApJ...737..103S}, respectively.
The process for constructing the model SEDs for the tracers is the same as mentioned in Section \ref{subsec:model_sed}, but we adopt $Z/Z_0 = 1/10$ for both NGC 6822 and WLM \citep{2012AJ....144..183L,2021ApJ...907...18R,2021arXiv211008793R}.
After result selection with the criteria mentioned in Section \ref{subsec:result_select}\footnote{Instead of $E(B-V) > 0.06$ mag, we consider the foreground extinction values for NGC 6822 and WLM as the second selection criterion listed in Section \ref{subsec:result_select}.}, 6 tracers in NGC 6822 and 4 tracers in WLM with reliable results are selected for further analysis.

\begin{figure}[ht!]
	\includegraphics[scale=0.1]{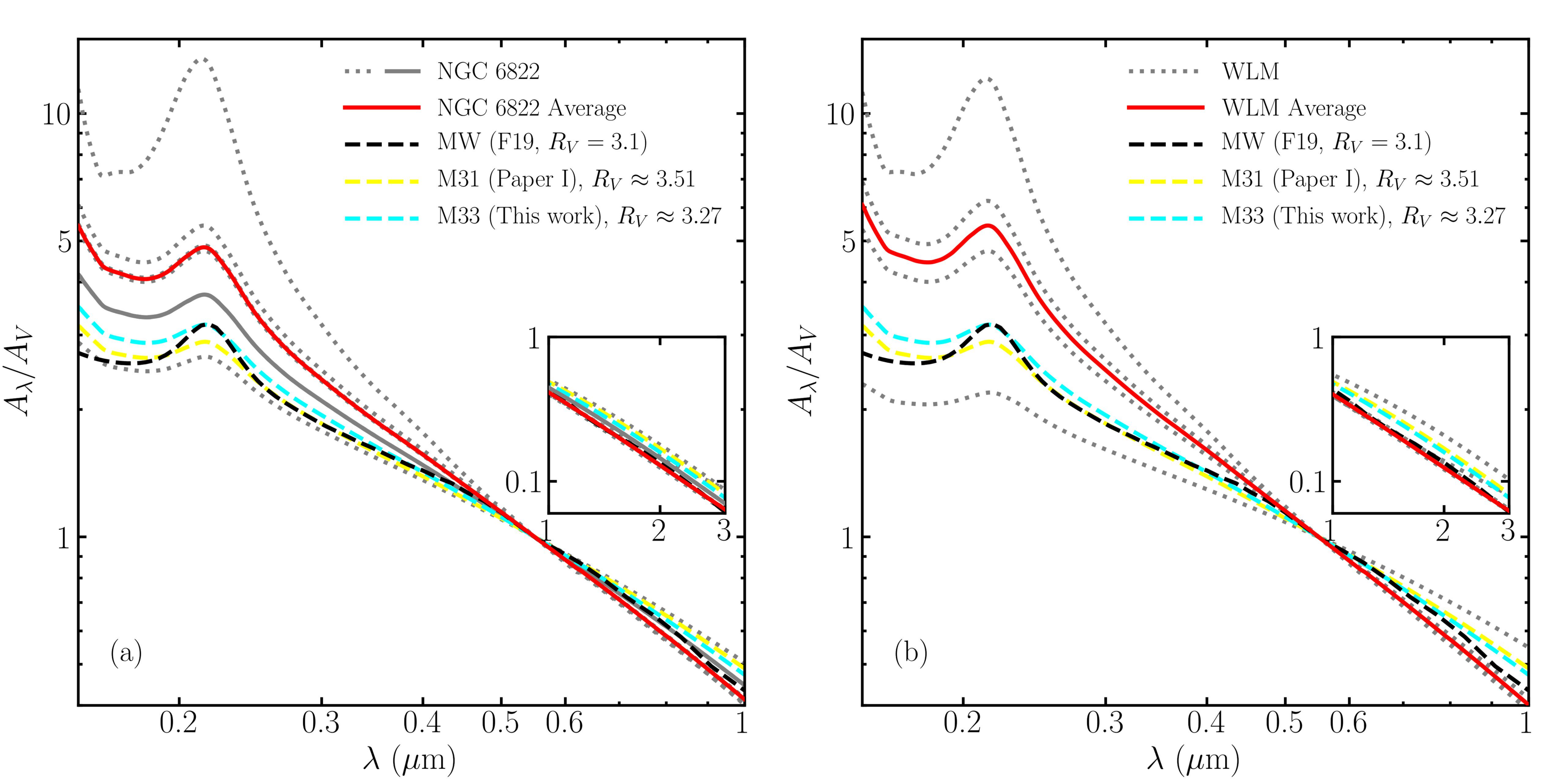}
	\caption{Dust extinction curves (gray dotted lines) toward the selected tracers in NGC 6822 [panel (a)] and in WLM [panel (b)].
	The solid gray line in panel (a) is the extinction curve toward the sight line with \emph{pcFlag = `UVI'} in NGC 6822.
	The solid red lines are the average ones for the derived extinction curves in NGC 6822 and WLM compared with those in the MW and M31 (the symbol convention follows Figure \ref{fig:M33ext}) as well as the average extinction curve in M33 derived in this work (cyan dashed line).
	\label{fig:LGGSext}}
\end{figure}

The results for all the selected tracers are listed in Table \ref{tab:re_LGGS}, and the derived extinction curves in NGC 6822 and WLM are plotted in Figure \ref{fig:LGGSext} (a) and (b), respectively, and compared with those in WM (F19, $R_V = 3.1$, black dashed line), M31 (\citetalias{2022ApJS..259...12W}, yellow dashed line) and M33 (this work, cyan dashed line).
There is only one extinction curve in NGC 6822 toward the sight line with \emph{pcFlag = `UVI'}, which shows a steeper far-UV rise than the average ones in the MW, M31 and M33, indicating a smaller dust size.
The average extinction curves presented in Figure \ref{fig:LGGSext} are calculated from the median values of derived $\alpha$ for the selected traces, which can only represent the average results in this work rather than the whole galaxies.
More photometric data in various bands are needed to obtain a better understanding of the extinction law and the dust properties in NGC 6822, WLM and other nearby star-resolved galaxies.

\movetabledown = 1.5in
\begin{rotatetable}
	\begin{deluxetable}{ccccccccccccccccc}
		\tablecaption{A list of the selected tracers in NGC 6822 and WLM with spectral type, fitting parameters ($\alpha,~{\rm log}~T_{\rm eff},~{\rm log}~g,~A_V$), derived $\overline{a}$, $E(B-V)$, \\ $R_V$ and information about the observed photometry adopted in this work$^a$.
		\label{tab:re_LGGS}}
		\tablehead{
		\colhead{Galaxy} & \colhead{LGGS ID} & \colhead{LGGS SpT} & \colhead{$\alpha$} & \colhead{${\rm log}(T_{\rm eff})$} & \colhead{${\rm log}(g)$} & \colhead{$A_V$} &\colhead{$\overline{a}$} & \colhead{$E(B-V)$} & \colhead{$R_V$} &  \colhead{Total bands} & \colhead{pcFlag}\\	
		\colhead{} & \colhead{} & \colhead{} & \colhead{} & \colhead{} & \colhead{}  & \colhead{(mag)} &  \colhead{(nm)} & \colhead{(mag)} &  \colhead{} & \colhead{}	& \colhead{ }
		}
		\startdata
		NGC 6822 & J194451.18-144919.8 & B0-1Ia & $3.26_{-0.04}^{+0.04}$ & $4.26_{-0.00}^{+0.00}$ & $3.00_{-0.08}^{+0.08}$ & $1.15_{-0.04}^{+0.04}$ & 7.93 & 0.30 & 3.77 & 10 & V \\
		NGC 6822 & J194455.47-144930.0 & B1-2I & $3.86_{-0.04}^{+0.17}$ & $4.18_{-0.01}^{+0.0}$ & $2.73_{-0.07}^{+0.08}$ & $0.74_{-0.06}^{+0.05}$ & 6.93 & 0.29 & 2.52 & 5 & V \\
		NGC 6822 & J194450.21-144253.6 & B1.5I & $3.89_{-0.07}^{+0.09}$ & $4.33_{-0.01}^{+0.02}$ & $2.73_{-0.07}^{+0.07}$ & $0.65_{-0.03}^{+0.04}$ & 6.89 & 0.26 & 2.48 & 5 & V \\
		NGC 6822 & J194452.28-145220.6 & B1I & $4.00_{-0.17}^{+0.07}$ & $4.19_{-0.01}^{+0.02}$ & $2.74_{-0.08}^{+0.07}$ & $0.95_{-0.04}^{+0.04}$ & 6.77 & 0.40 & 2.35 & 8 & VI \\
		NGC 6822 & J194455.08-145213.1 & B5Ia & $5.65_{-0.09}^{+0.06}$ & $4.20_{-0.00}^{+0.05}$ & $2.49_{-0.08}^{+0.09}$ & $1.27_{-0.05}^{+0.07}$ & 5.83 & 0.65 & 1.96 & 4 & V \\
		NGC 6822 & J194501.60-145440.0 & B8-A0:I & $3.62_{-0.08}^{+0.06}$ & $4.16_{-0.03}^{+0.03}$ & $2.25_{-0.09}^{+0.09}$ & $0.82_{-0.17}^{+0.05}$ & 7.26 & 0.28 & 2.91 & 9 & UVI \\
		NGC 6822 & Average$^b$ & - & 3.88 & - & - & 0.89 & 6.91 & 0.30 & 2.50 & - & - \\
		WLM & J000157.20-152718.0 & B1.5Ia & $4.14_{-0.24}^{+0.04}$ & $4.43_{-0.00}^{+0.00}$ & $2.74_{-0.08}^{+0.07}$ & $0.46_{-0.04}^{+0.04}$ & 6.64 & 0.21 & 2.21 & 5 & V \\
		WLM & J000158.12-152648.5 & B1.5Ia & $3.86_{-0.41}^{+1.53}$ & $4.31_{-0.02}^{+0.04}$ & $2.73_{-0.07}^{+0.07}$ & $0.25_{-0.04}^{+0.03}$ & 6.92 & 0.10 & 2.52 & 5 & V \\
		WLM & J000153.22-152839.5 & B9Ia & $3.03_{-0.09}^{+0.07}$ & $4.13_{-0.02}^{+0.01}$ & $2.26_{-0.09}^{+0.09}$ & $0.45_{-0.05}^{+0.04}$ & 8.56 & 0.10 & 4.57 & 12 & VI \\
		WLM & J000159.95-152819.0 & O9.7Ia & $5.17_{-0.46}^{+0.5}$ & $4.45_{-0.01}^{+0.01}$ & $3.25_{-0.09}^{+0.09}$ & $0.25_{-0.03}^{+0.04}$ & 6.00 & 0.13 & 1.90 & 12 & VI\\
		WLM & Average$^b$ & - & 4.00 & - & - & 0.35 & 6.78 & 0.12 & 2.37 & - & - \\
        \enddata	
		\tablecomments{$^a$ The columns descriptions are the same as those for Table \ref{tab:re}.\\
		$^b$ These lines show the median values of the derived parameters for selected tracers in NGC 6822 and WLM.
		}
	\end{deluxetable}
\end{rotatetable}

% ==================

\section{Summary} \label{sec:summary}

A sample of bright O-type and B-type supergiants from the LGGS catalog \citep{2016AJ....152...62M} are chosen as extinction tracers to derive the dust extinction curves in M33.
This is the first study focused on the dust extinction curves toward individual sight lines in M33 rather than the dust attenuation curves
in \citet{1999ApJ...519..165G,2017PhDT.......221H}.
The previous studies have been improved, and the main results of this work are as follows:

1. The extinction curves in M33 derived in this work cover a wide range of shapes, from curves with an obvious 2175 $\,{\rm \AA}$ bump (like the extinction curves with $R_V \approx 2$) to relatively flat curves with $R_V \approx 6$, implying the complexity of the interstellar environment and the inhomogeneous distribution of interstellar dust in M33.
The derived parameter $\alpha$ in the dust size distribution ranges from $\approx 2.6-5.9$, while the dust size ranges from $\approx 5.78 - 9.64$ nm.

2. The average extinction curve in M33 ($R_V \approx 3.39$) is similar to the MW extinction curve with $R_V = 3.1$ but with a slightly weaker 2175 $\,{\rm \AA}$ bump and a slightly steeper far-UV rise.
The average dust size distribution in M33 is $dn/da \sim a^{-3.45}{\rm exp}(-a/0.25)$, and the median value of the average dust size is $\overline{a} \approx 7.54$ nm, which is smaller than that of the MW ($\overline{a_{\rm MW}} \approx 8.45$ nm).

3. The derived $A_V$ in M33 is up to 2 mag with a median value of $\approx 0.43$ mag, which is smaller than the median value ($A_V \approx 0.53$ mag) derived in \citet{2017PhDT.......221H} and larger than the mean amount ($A_V \approx 0.25$ mag) measured by \citet{2009A&A...493..453V}.

4. The method adopted in this work and \citetalias{2022ApJS..259...12W} that combines the stellar model atmospheres and the dust models to calculate extinction curves and analyze dust properties toward individual sight lines is extended to the star-resolved galaxies NGC 6822 and WLM, but we can only derive the extinction curves toward a few individual sight lines.
More observations are needed to gain a better understanding of the extinction law and dust properties in nearby star-resolved galaxies.

% ==========

\acknowledgments
We greatly thank the anonymous reviewer for the very helpful suggestions, which improved the paper significantly.
It is a pleasure to thank Prof. Biwei Jiang, Prof. Haibo Yuan, Dr. Qi Li and Dr. Shu Wang for the very helpful discussions.
This work is supported by the National Natural Science Foundation of China through projects NSFC 12133002 and U2031209 and the CSST Milky Way and Nearby Galaxies Survey on Dust and Extinction Project CMS-CSST-2021-A09.






\bibliography{paper}{}
\bibliographystyle{aasjournal}



\end{CJK*}
\end{document}